\documentclass[fleqn,usenatbib]{mnras}
\usepackage[dvipsnames]{xcolor}
\usepackage{siunitx}
\usepackage{graphicx}	
\usepackage{amsmath}	
\usepackage{amssymb}
\usepackage{hyperref}
\usepackage{tabularx}
\usepackage{mathptmx}
\usepackage{txfonts}
\usepackage{academicons}
\definecolor{orcidlogocol}{HTML}{A6CE39}
\usepackage[T1]{fontenc}
\usepackage{longtable}
\usepackage{subcaption}
\usepackage[symbol]{footmisc}
\usepackage{multicol, blindtext}
\usepackage[justification=centering]{caption}
\providecommand{\bjdtdb}{\ensuremath{\rm {BJD_{TDB}}}}
\providecommand{\feh}{\ensuremath{\left[{\rm Fe}/{\rm H}\right]}}
\providecommand{\teff}{\ensuremath{T_{\rm eff}}}
\providecommand{\ecosw}{\ensuremath{e\cos{\omega_*}}}
\providecommand{\esinw}{\ensuremath{e\sin{\omega_*}}}
\providecommand{\msun}{\ensuremath{M}}
\providecommand{\rsun}{\ensuremath{R}}
\providecommand{\lsun}{\ensuremath{L}}
\providecommand{\mj}{\ensuremath{\,M_{\rm J}}}
\providecommand{\rj}{\ensuremath{\,R_{\rm J}}}
\providecommand{\mpp}{\ensuremath{\,M_{\rm P}}}
\providecommand{\rp}{\ensuremath{\,R_{\rm P}}}

\providecommand{\fave}{\langle F \rangle} 
\providecommand{\fluxcgs}{10$^9$ erg s$^{-1}$ cm$^{-2}$}

\providecommand{\arcsec}{$^{\prime \prime}$}
\providecommand{\arcmin}{$^{\prime}$}

\DeclareRobustCommand{\VAN}[3]{#2}
\let\VANthebibliography\thebibliography
\def\thebibliography{\DeclareRobustCommand{\VAN}[3]{##3}\VANthebibliography}
\newcommand{\orcid}[1]{\href{https://orcid.org/#1}{\includegraphics[width=8pt]{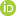}}}

\title{Discovery of an inflated hot Jupiter around a slightly evolved star TOI-1789}

\author[Akanksha Khandelwal]{{
Akanksha Khandelwal{\thanks{E-mail: akankshak@prl.res.in}}$^{1,2}$ \orcid{0000-0003-0335-6435},
Priyanka Chaturvedi$^{3}$ \orcid{0000-0002-1887-1192},
Abhijit Chakraborty$^{1}$\orcid{0000-0002-3815-8407},
Rishikesh Sharma$^{1}$\orcid{0000-0001-8983-5300}}, 
\newauthor{Eike. W. Guenther$^{3}$,
Carina M. Persson$^{4}$,
Malcolm Fridlund$^{4,5}$\orcid{0000-0003-2180-9936},
Artie P. Hatzes$^{3}$,
Neelam J.S.S.V. Prasad$^{1}$\orcid{0000-0003-0670-5821},
}
\newauthor{Massimiliano Esposito$^{3}$, 
Sireesha Chamarthi$^{3,6}$,
Ashirbad Nayak$^{1}$,
Dishendra$^{1}$\orcid{0000-0003-3747-7809},
Steve~B.~Howell$^{7}$\orcid{0000-0002-2532-2853}
}
\\\\
$^{1}$ Astronomy $\&$ Astrophysics Division, Physical Research Laboratory, Ahmedabad 380009, India\\
$^{2}$Indian Institute of Techonology, 382355 Gandhinagar, India\\
$^{3}$Thüringer Landessternwarte Tautenburg, Sternwarte 5, 07778 Tautenburg, Germany\\
$^{4}$Department of Space, Earth and Environment, Chalmers University of Technology, Onsala Space Observatory, 439 92 Onsala, Sweden \\
$^{5}$Leiden Observatory, University of Leiden, PO Box 9513, 2300 RA, Leiden, The Netherlands \\
$^{6}$ Deutsches Zentrum für Luft- und Raumfahrt-Institut für Datenwissenschaften, Mälzerstraße 3, 07745 Jena, Germany\\
$^{7}$ NASA Ames Research Center, Moffett Field, CA 94035, USA
}

\date{Accepted 2021 October 8. Received 2021 October 8; in original form 2021 June 15.}
\pubyear{2021}

\begin{document}
\label{firstpage}
\pagerange{\pageref{firstpage}--\pageref{lastpage}}
\maketitle

\begin{abstract}
We report here the discovery of a hot Jupiter at an orbital period of $3.208664\pm0.000015$ days around TOI-1789 (TYC 1962-00303-1, $TESS_{mag}$ = 9.1) based on the TESS photometry, ground-based photometry, and high-precision radial velocity observations. The high-precision radial velocity observations were obtained from the high-resolution spectrographs, PARAS at Physical Research Laboratory (PRL), India, and TCES at Thüringer Landessternwarte Tautenburg (TLS), Germany, and the ground-based transit observations were obtained using the 0.43~m telescope at PRL with the Bessel-$R$ filter. The host star is a slightly evolved ($\log{g_*}$ = $3.943^{+0.023}_{-0.043}$), late F-type (\teff = $5991\pm55$ K), metal-rich star (\feh = $0.373^{+0.071}_{-0.086}$ dex) with a radius of {\ensuremath{\,R_*}} = $2.168^{+0.036}_{-0.034}$ \(R_\odot\)  located at a distance of $223.53^{+0.91}_{-0.90}$ pc. The simultaneous fitting of the multiple light curves and the radial velocity data of TOI-1789 reveals that TOI-1789~b has a mass of $\mpp$ = $0.70\pm0.16 $ $\mj$, a radius of $\rp$ = $1.44^{+0.24}_{-0.14}$ $\rj$, and a bulk density of $\rho_P$ = $0.28^{+0.14}_{-0.12}$ g cm$^{-3}$ with an orbital separation of a = $0.04882^{+0.00063}_{-0.0016}$ AU. This puts TOI-1789~b in the category of inflated hot Jupiters. It is one of the few nearby evolved stars with a close-in planet. The detection of such systems will contribute to our understanding of mechanisms responsible for inflation in hot Jupiters and also provide an opportunity to understand the evolution of planets around stars leaving the main sequence branch.
\end{abstract}


\section{Introduction}\label{sec:intro}
Hot Jupiters have set several firsts, from being the first kind of exoplanets discovered with the Radial Velocity (RV) method \citep[51 Peg;][]{1995Natur.378..355M}, to being also the first kind of transiting systems studied for precise radii and orbits \citep[HD 209458 b;][]{2000ApJ...529L..45C, 2000ApJ...529L..41H}. These systems have intrigued the theorists triggering a critical revision of planet formation theories. Two theories have been used the most to explain close-in giant planets. These are the in-situ formation by gravitational instability \citep{1997Sci...276.1836B,2007prpl.conf..607D} or core accretion followed by inward migration \citep{1974Icar...22..416P,2014prpl.conf..619C}. Gravitational instability envisions the formation of giant planets via the fragmentation of proto-planetary disk into bound clumps \citep{1997Sci...276.1836B}. In the case of core accretion, the giant planet is formed at several AUs beyond the so-called ice-line where sufficient solid material is present to form the core, and then it migrates inward. The migration is caused by either torques from the proto-planetary disk or by  gravitational scattering due to a third body. This can shrink the orbital separation of the planet from their formation location of several AUs to hundredths of AUs \citep{1986ApJ...309..846L,high_e_migration,2014prpl.conf..667B}. One may refer \cite{lower_mss_limit} for a detailed review on the formation mechanism of hot Jupiters. Although a large number of close-in giant planets have been detected, this is largely a bias effect due to the ease in detecting them (high RV amplitude, short periods). The true frequency of close-in planets is only $\sim$ 1$\%$ (\cite{2010PASP..122..905J,2011arXiv1109.2497M,2019AJ....158..109H} and references therein).

Space-based transit surveys like CoRoT \citep{2006cosp...36.3749B}, Kepler \citep{KEPLER} and TESS \citep{Ricker2015} have not only increased the number of discovered hot Jupiters, but their high photometric precision has allowed for the determination of the most precise planet radii and densities. These surveys have revealed that hot Jupiters tend to have inflated radii, typically $10-50\%$ larger than that of Jupiter \citep{2012AJ....144..139H, hats26, 2018MNRAS.478.5356S, 2018MNRAS.481.4960R, 2021MNRAS.tmp..893T}. In some of the systems, heat is known to slow down the contraction of the planet \citep{2002A&A...385..166S,2017ApJ...844...94K}. Several mechanisms have been proposed to explain this inflation, one being star-planet tidal interactions \citep{2003ApJ...592..555B} and the other in which kinetic energy from the wind is converted into heat \citep{2002A&A...385..156G} and Ohmic dissipation \citep{2010ApJ...714L.238B}. The most likely mechanism responsible for such inflation is the large incident stellar irradiation \citep{2011ApJ...729L...7L, 2018AJ....155..214T}. However, this process alone is not enough to explain the inflation. It was suggested by \cite{2002A&A...385..166S} that energy is dissipated by atmosphere circulation due to the transport of kinetic energy to the interior of planets which could additionally be contributing to the process of inflation \citep{inf}. Evolved stars provide higher source of radiation due to their larger sizes compared to their solar analogues. There are only a handful of systems with giant stars harbouring planets interior to 0.5 AU (eg. \citealt{2014A&A...568L...1L, 2015A&A...573A...3J, 2015A&A...573L...6O}).
Thus, it is imperative to look at stars which might be leaving the main sequence phase and entering the sub-giant/giant branch. It is interesting to catch these stars in an evolutionary phase that has yet not caused the engulfment of its orbiting planet. There have been only a handful of such close-in planets discovered as of now orbiting around slightly evolved stars, e.g., KELT-12~b \citep{2017AJ....153..178S} and TOI-640~b \citep{2021AJ....161..194R} to name a few.

In this paper, we present the discovery of a hot Jupiter around a star TYC 1962-00303-1 or TESS Object of Interest TOI-1789, an important contribution to this field of relatively rare kinds of planets around slightly evolved stars. We use spectroscopic, photometric, and imaging data to confirm the nature of this transiting candidate and to characterize the planet in terms of its mass, radius and density. All the observations are described in Section~\ref{sec:obs} of the paper. In Section~\ref{sec:analysis}, we focus on the host star and characterization of the star-planet system in detail with the joint modeling of transit photometry and RVs. In Section~\ref{sec:discussion}, we discuss the implications of our results with summarizing the paper in Section~\ref{sec:summary}.

\section{Observations}\label{sec:obs}
\subsection{TESS Photometry}
TYC~1962-00303-1 is a relatively bright star ($V$ = 9.7 mag) in the northern celestial hemisphere, first listed as TOI-1789 on March 12, 2020. This source was observed by TESS between Jan 21 and Feb 18, 2020 (27.3 days time span) with a gap of ${\sim} $2 days due to the data transferring from the spacecraft. TESS operates at a wavelength region of  600 to 1000 nm and covers 21\arcsec\space of sky in each pixel \citep{Ricker2015}. It has four identical cameras with a 24$^{\circ}\times24^{\circ}$ field of view for each camera. TOI-1789 was observed in the CCD-4 of camera 1 in sector 21 in a long cadence (30 minutes) mode. 

The data were analyzed with the Quick Look Pipeline \citep[QLP:][]{2020RNAAS...4..204H} developed by MIT and the Science Processing Operations Center \citep[SPOC:][]{spoc} pipeline, based on the Kepler mission pipeline at the NASA Ames research center. Both pipelines detected eight transits with a depth of $\sim$ 2600 ppm for TOI-1789 spaced at a period of $\sim$ 3.21 days and a duration of $\sim$ 2.3 hours. The transit data is also vetted by the TESS Science team before releasing it as a planetary candidate. The transits are reported to be V-shaped and a slight centroid offset in the target position. We adopt the Pre-search Data Conditioning Simple Aperture Photometry (PDCSAP) light curves \citep{Stumpe_2014,2012PASP..124.1000S}, provided by the SPOC pipeline, which is publically available at the Mikulski Archive for Space  Telescopes (MAST)\footnote{\url{https://mast.stsci.edu/portal/Mashup/Clients/Mast/Portal.html}}. We applied the Box Least-Square \citep[BLS;][]{box_least} periodogram to detect the transits in the TESS light curve using PDCSAP fluxes. We successfully retrieved the transit signal at a period of 3.2076208 with a depth of 2610$\pm$220 ppm. 
To search for additional transit signals we removed this 3.2076208 days signal from the data and re-ran the BLS periodogram. We did not find any other significant peaks. For further analysis, we use median-normalized PDCSAP fluxes that are additionally detrended by fitting a high-order polynomial over out-of-transit data using the \texttt{lightkurve} package \citep{lightkurve}.
The basic properties of the star found in the literature are listed in Table~\ref{tab:star_table}.
\begin{table}
	\begin{center}

	\caption{Basic Stellar Parameters for TOI-1789}
	\label{tab:star_table}
	\begin{tabular}{cllc} 
		\hline
		\hline
		Parameter & Description (unit) & Value & Source\\
		\hline
		$\alpha_{J2000}$ & Right Ascension & 09 30 58.42 & (1)\\
		$\delta_{J2000}$ & Declination & 26 32 23.98 & (1)\\
		$\mu_{\alpha}$ & PM in R.A. (mas yr$^{-1}$) & -7.977 $\pm$ 0.019 & (1)\\
		$\mu_{\delta}$ & PM in Dec (mas yr$^{-1}$) & -39.401 $\pm$ 0.015 & (1)\\
		$\pi$ & Parallax (mas) & 4.474 $\pm$ 0.0181 & (1)\\
        $G$ & $Gaia$ G mag & 9.584 $\pm$ 0.0002 & (1)\\
		$T$ & TESS T mag & 9.182 $\pm$ 0.006 & (2)\\
		$B_{T}$ & Tycho B mag  &10.422 $\pm$ 0.039 & (3)\\
		$V_{T}$ & Tycho V mag & 9.788	 $\pm$ 0.031 & (3)\\
		$B$   & APASS B-mag & 10.335$\pm$ 0.020 & (6)\\
        $V$   & APASS V-mag  &9.686 $\pm$0.030 & (6)\\
        $g$   & SDSSg mag  &10.353$\pm$ 0.100 & (6)\\
        $r$   & SDSSr mag  &9.590$\pm$ 0.060 & (6)\\
        $i$   & SDSSi mag  &9.398$\pm$ 0.020 & (6)\\
		$J$   & 2MASS J mag & 8.672 $\pm$ 0.024 & (4)\\
		$H$   & 2MASS H mag & 8.410 $\pm$ 0.021 & (4)\\
		$K_{S}$ & 2MASS K$_S$ mag & 8.345 $\pm$ 0.018 & (4)\\
		$W1$  & WISE1 mag & 8.297 $\pm$ 0.023 & (5)\\
		$W2$  & WISE2 mag & 8.348 $\pm$ 0.02 & (5)\\
		$W3$  & WISE3 mag & 8.311 $\pm$ 0.024 & (5)\\
		$W4$  & WISE4 mag & 7.996 $\pm$ 0.226 & (5)\\
	
	\hline
		
	\end{tabular}
		\end{center}
	Other Identifiers:
	\begin{center}
	    HD 82139$^7$\\
	    TIC 172518755$^2$\\
	    TYC 1962-00303-1$^3$\\
	    2MASS J09305841+2632246$^4$\\
	    $Gaia EDR3$ 646125297938578944$^1$\\

	\end{center}
	\smallskip
	\hrule
	\smallskip
	\textbf{Note:} The spectral type for TOI-1789 reported by SIMBAD is K0, however with our spectral analysis we find it to be a late F-type star (see Section~\ref{sec:sme} and Section~\ref{sec:evolved}).\\
	\textbf{References.} (1) \cite{gaiaedr3}, (2) \cite{2018AJ....156..102S}, (3) \cite{tycho}, (4) \cite{JHK}, (5) \cite{ALLWISE}, (6) \cite{APASS}, (7) \cite{hd_id}\\
	
\end{table}

\begin{figure*}
    \begin{subfigure}{.5\textwidth}
        \centering
            \includegraphics[width=0.72\columnwidth]{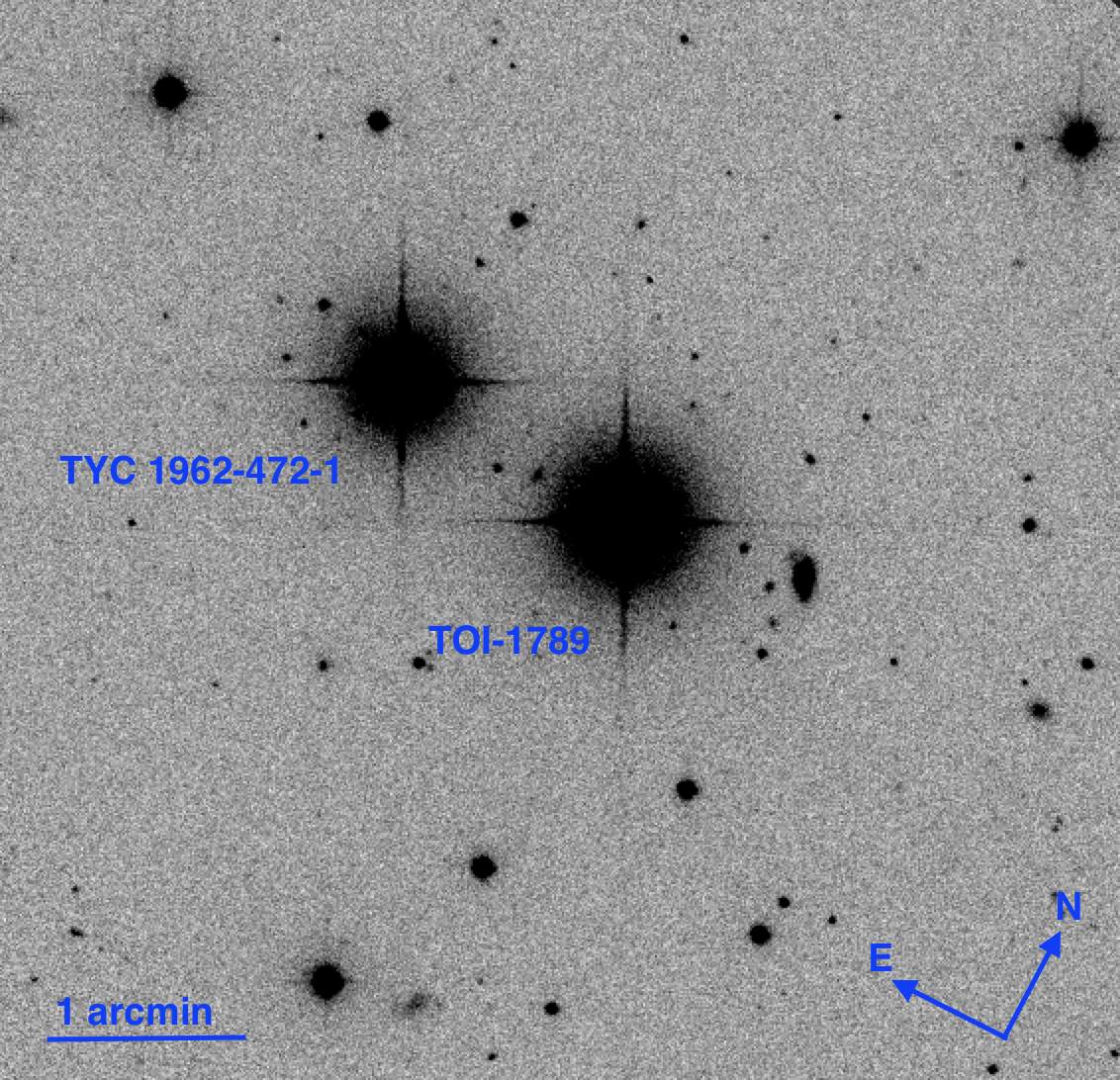}
        \label{fig:sub1}
    \end{subfigure}%
    \begin{subfigure}{.5\textwidth}
        \centering
            \includegraphics[width=\columnwidth]{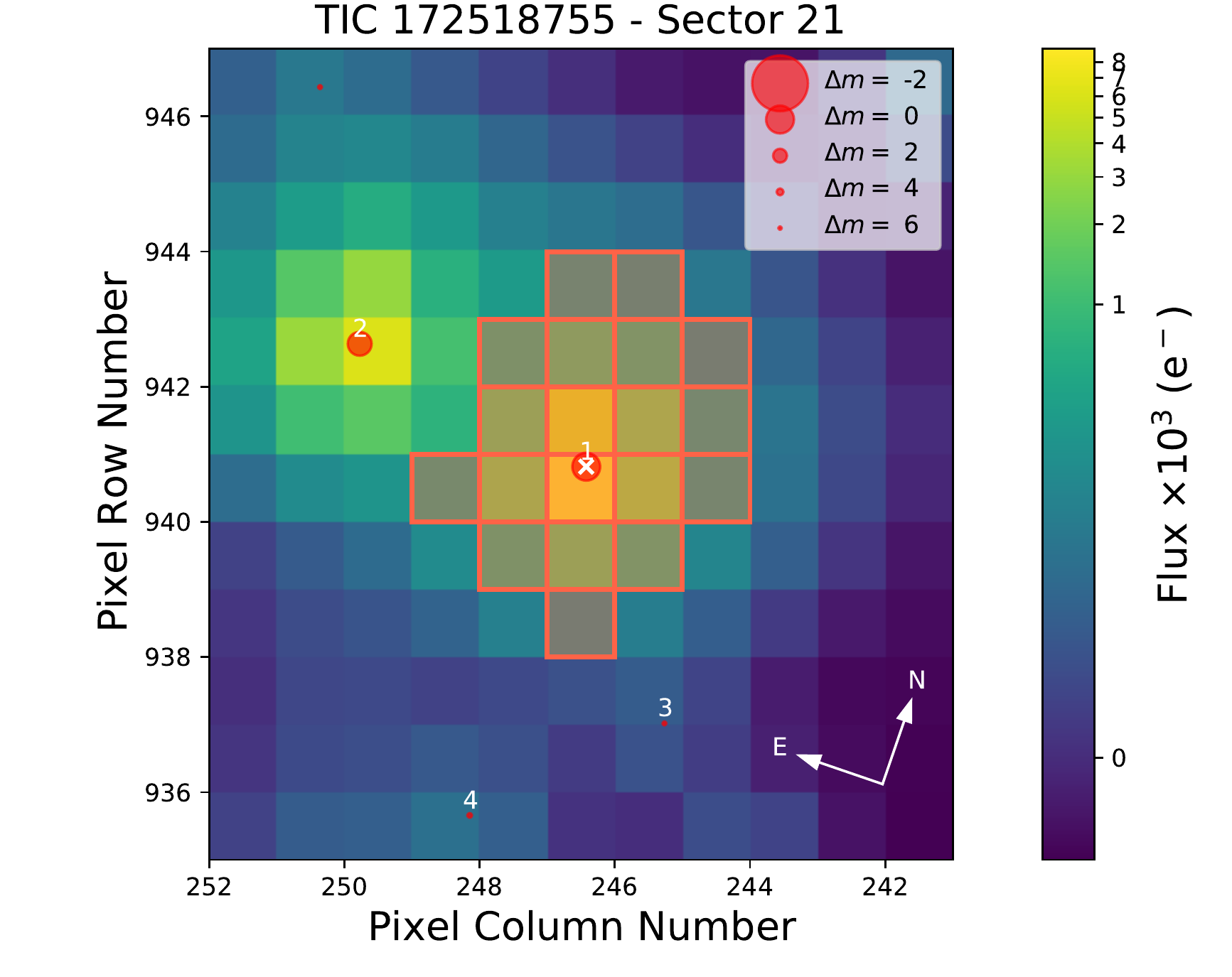}
        \label{fig:sub2}
    \end{subfigure}
    \caption{\textit{Left panel:} The SDSS (DR7) image of TOI-1789 and its visual binary companion (TYC 1962-472-1) in \textit{sloan-i} band{\protect\footnotemark}. \textit{Right panel:} Target pixel file (TPF) of TOI-1789 in TESS sector 21 generated with \texttt{tpfplotter} \citep{tpfplot}. The sizes of individual red dots represent the magnitude contrast ($\Delta$\textit{m}) with TOI-1789. The red-squared region represents the aperture used for photometry by SPOC and the position of TOI-1789 and TYC 1962-472-1 is labelled as `1' and `2', respectively.}
    \label{fig:tpfplot}
\end{figure*}

The system is reported to be a widely separated visual binary with the secondary star TYC 1962-475-1 of similar spectral type and brightness (see Figure~\ref{fig:tpfplot}), and with an orbital separation of 17776 AU (1.3\arcmin separation on the sky)  \citep{2017MNRAS.472..675A}. The Renormalised Unit Weight Error (RUWE) number from the $Gaia$ catalog for the target star is 0.948, a value $\sim$ 1, indicative that a single-star model is used as the best fit for astrometric observations \citep{gaiaedr3}. RUWE > 1.4 could indicate a binary nature for the astrometric solution used in $Gaia$. The on-sky separation for these two sources is larger than the TESS pixel scale of 21\arcsec.

\footnotetext{https://irsa.ipac.caltech.edu/applications/finderchart/}
Based on the {Kepler} manual \citep{2016ksci.rept....9T}, these PDCSAP fluxes are corrected for any contamination from neighboring pixels. However, due to the V-shape, short transit duration ($\sim$~2 hrs), and the long cadence data (30 min), the transit shape, transit duration, and the planetary radius can not be determined precisely. There is also a small centroid offset in the target's position given in the TESS data validation report. We, therefore, conducted short-cadence ($\sim$~20s) ground-based transit observations to precisely obtain the transit parameters and to understand and verify the nature of the transits detected by TESS (see Section~\ref{sec:prl_photo}).

\subsection{Ground-based Photometry}\label{sec:prl_photo}
\begin{figure}
    \centering
	\includegraphics[width=15cm, height=15cm, trim={0 2cm 0 0cm}]{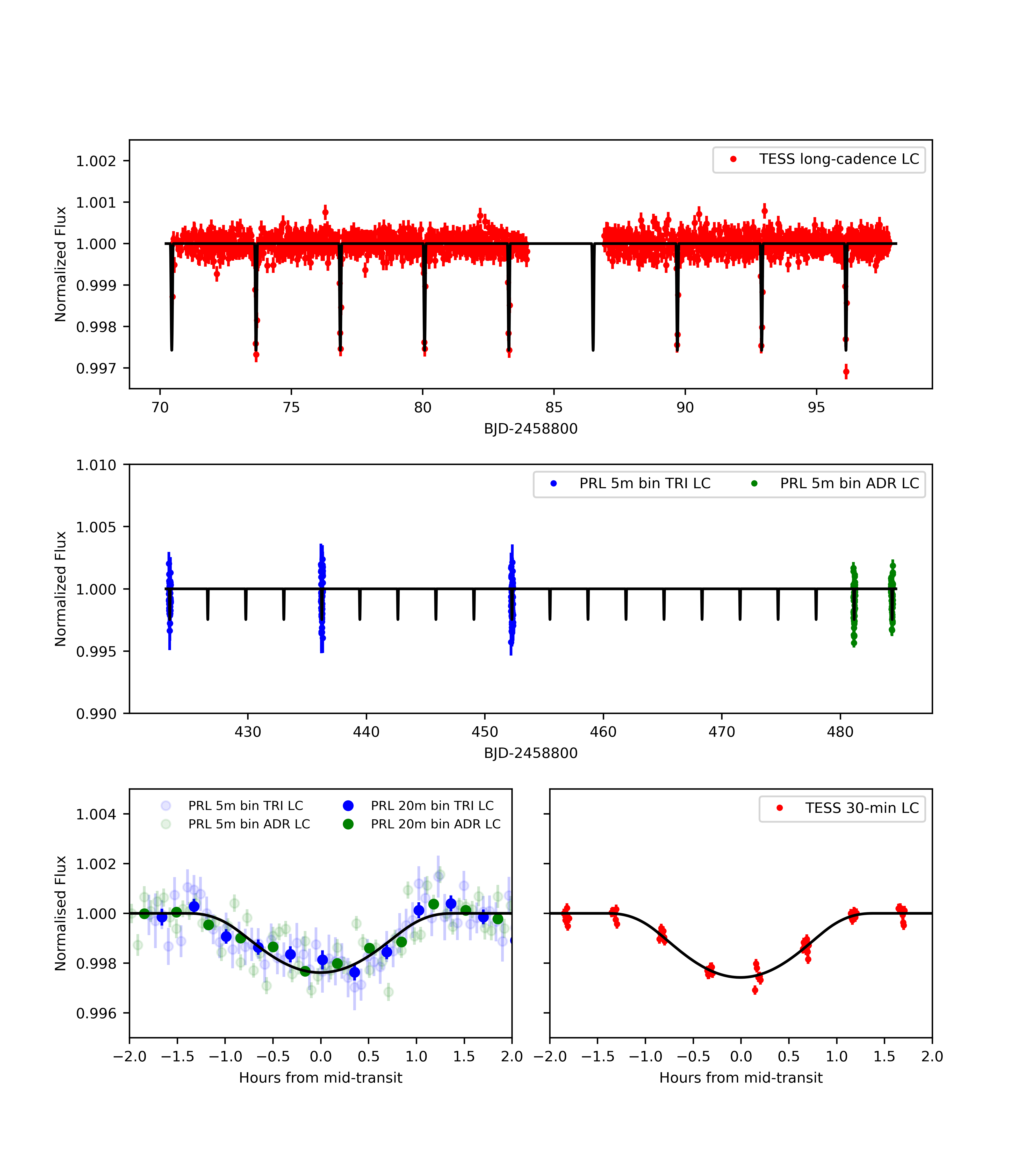} 
	\smallskip\\
    \caption{Upper panel:- The normalized TESS Light Curve (LC) of TOI-1789 is plotted in red. Eight transits can be seen spaced at ${\sim}3.21$ days with a depth of ${\sim}2.6$ ppt. Middle Panel:- The ground-based  follow-up photometry with PRL 0.43~m telescope is shown here (Section~\ref{sec:prl_photo}). The blue dots represent the three transits observed with a {\tt{TRIUS}} CCD (TRI) with a precision of 1.16 ppt ($\sim$ 1.3 mmag), while green dots represent the two transit events observed with an {\tt{ANDOR}} CCD (ADR) with a precision of 0.92 ppt ($\sim$ 1 mmag) in 5-min bins. Lower-left panel:- All the transits identified by the CCD detector are phased to the orbital period of TOI-1789~b and then binned to 5-min and 20-min cadence are plotted here, with blue and green color depicting the TRI and ADR datasets, respectively.
    Lower-right panel:- Red dots represent the phase folded TESS LC (30-min cadence). Note:- In all the panels, the over-plotted black line represents the best-fit transit model from EXOFASTv2, obtained by simultaneous fitting of TESS and PRL photometry data. (For details, refer Section~\ref{sec:exofast}) }
    \label{fig:TESS_light_curve}
\end{figure}

We acquired ground-based transit follow-up observations for TOI-1789 with the 0.43~m telescope ({\tt {CDK17 from PLANEWAVE}})\footnote{https://planewave.com/product/cdk17-ota/} located at {Gurushikhar Observatory, PRL, Mt. Abu, India}. Five transits were observed between January 08, 2021 and March 10, 2021 as summarised in Table~\ref{tab:photo_table}.
  The first three transits were observed using a low-cost {\tt {TRIUS PRO-814}} (TRI) CCD\footnote{https://www.sxccd.com/product/trius-sx814/} which provides a field of view (FOV) of 14.6\arcmin $\times$ 11.7\arcmin\space with a pixel scale of 0.26\arcsec. For the last two transits an {\tt {ANDOR iKon-L 936}} (ADR) CCD\footnote{https://andor.oxinst.com/products/ikon-xl-and-ikon-large-ccd-series/ikon-l-936} was used, which has an FOV of 32\arcmin $\times$ 32\arcmin with a pixel scale of 0.95\arcsec. The telescope was slightly defocused while observing these events to have longer exposure times, which resulted in high SNR and thus, helped in achieving higher photometric precision. All the transits were observed in the Bessel-$R$ passband and the main specifications of both the CCDs are listed in Table~\ref{tab:ccds}. 
  The use of {\tt {ANDOR iKon-L 936}} CCD in precision differential photometric observations is well established as the same kind of CCD detectors are used in various transit surveys like the SuperWASP \citep{superwasp} and the SPECULOOS \citep{speculoos}.
We used the AstroImageJ software (AIJ: \cite{2017AJ....153...77C}) for data reduction and light curve extraction. AIJ is an efficient tool for time-series ultra-precise differential photometry (e.g., exoplanet transits), light curve detrending, fitting, and plotting. First, the raw science frames were dark-corrected and flat fielded. Then, for the lightcurve extraction we did the multi-aperture differential photometry as described in the \cite{2017AJ....153...77C}. An aperture of 1.5 times the FWHM of the star is used. 
The extracted light curves for all the five transit observations were detrended with respect to FWHM, airmass, and exposure time. These parameters are linearly detrended and their contribution is added to the overall $\chi^2$ contribution for the curve fitting within AIJ for each transit (see equation 5 in \cite{2017AJ....153...77C}. 
Finally, the normalized light curves were extracted from AIJ. The transit event was clearly visible in all the detrended LCs, as shown in Figure~\ref{fig:TESS_light_curve}. For better display purpose, we have binned the LC to 20 min. 
All the transits were jointly modelled with the observed RV data (Section~\ref{sec:exofast}).
The residuals from the best-fit transit model have a standard deviation of 1.16 ppt (parts per thousand, ($\sim$ 1.3 mmag)) and 0.92 ppt ($\sim$ 1 mmag) in 5 min bins for TRI and ADR datasets, respectively. The precision achieved with {\tt {TRIUS}} CCD camera is within the 1.3~$\sigma$ of the precision achieved with {\tt {ANDOR}} CCD camera showing that the low-cost CCD camera like {\tt {TRIUS PRO-814}} can also be used for precision differential photometric observations effectively.
\onecolumn
\begin{longtable}{ccccccc}
\caption{A summary of the ground-based transit follow-up observations}
\label{tab:photo_table}\\ \hline
		\noalign{\smallskip}
	    Instrument & UT Date & Coverage & 5-min Precision (ppt) & Avg. PSF (\arcsec) & Avg. EXP-TIME\\
	    \noalign{\smallskip}
		\hline
		\noalign{\smallskip}
		\tt{TRIUS PRO-814} & 08 Jan 2021 & Full & 0.94 & 4.1 & 25s\\
		\tt{TRIUS PRO-814} & 21 Jan 2021 & Full & 1.04 & 4.4 & 25s\\
		\tt{TRIUS PRO-814} & 06 Feb 2021 & Full & 1.32 & 3.9 & 20s\\
		\tt{ANDOR iKon-L 936}  & 07 Mar 2021 & Full & 0.89 & 6.2 & 18s\\
		\tt{ANDOR iKon-L 936}  & 10 Mar 2021 & Full & 1.00 & 4.5 & 8s\\
		\noalign{\smallskip}
		\hline
\end{longtable}

\twocolumn

\begin{table}
\centering
\caption{Specifications of both the CCDs}
\label{tab:ccds}
\begin{tabular}{lcc}
\hline
\noalign{\smallskip}
		Specifications &  {\tt {TRIUS PRO-814}} & {\tt {ANDOR iKon-L 936}}  \\
		\noalign{\smallskip}
		\hline
		\noalign{\smallskip}
		Manufacturer & Starlight Xpress Ltd. & Oxford Instruments \\
		Sensor type & Monochrome ICX814AL & e2V CCD42-40 \\
		 & (Interline CCD) & (BEX2-DD) \\
		Image format (pix)& 3388 x 2712 & 2048 x 2048 \\
		Pixel size ($\mu$m) & 3.69 & 13.5 \\
		QE (at 580~nm) & $\sim$ 77$\%$  &  $\ge90\%$ \\
		Dark current (e$^{-}$/s/pix)& $\leq$ 0.002 at \SI{-10}{\celsius} &  0.006 at \SI{-80}{\celsius}\\
		System gain (e$^{-}$/ADU) & 0.25 & 1 \\
		Readnoise & 3e$^{-}$ at 3~MHz& 7e$^{-}$ at 1~MHz \\
		\noalign{\smallskip}
		\hline
\end{tabular}
\end{table}
 
\subsection{High-resolution Speckle Imaging}

\begin{figure}
	\includegraphics[width=\columnwidth]{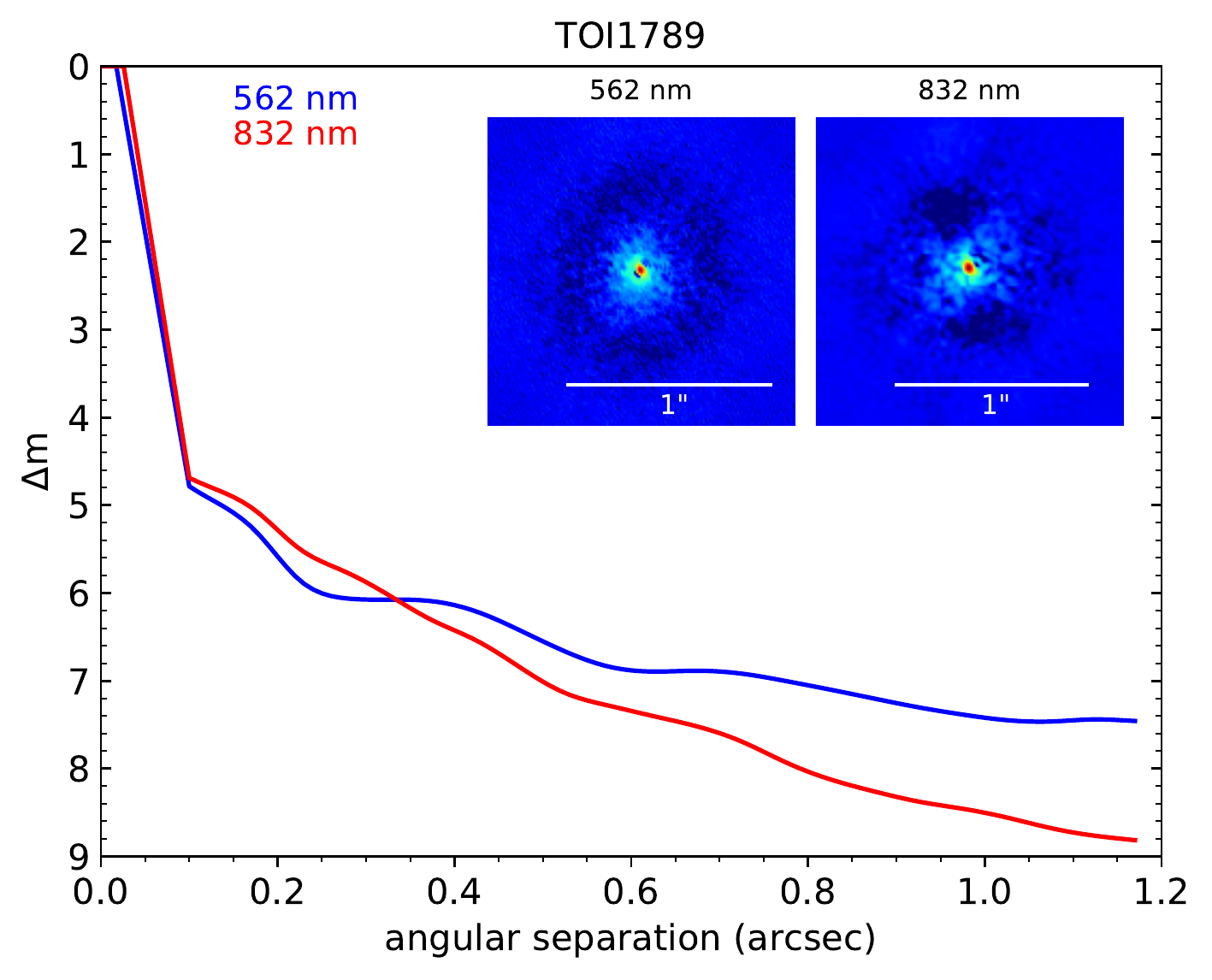}
    \caption{ Gemini observations of TOI-1789 on UT 3 February 2021. The 5$\sigma$ contrast curves are plotted with their reconstructed images.}
    \label{fig:Gemini_speckle_curve}
\end{figure}

High resolution imaging is a tremendously useful tool for finding contamination from nearby star. TOI-1789 was observed on February 3, 2021 UT using the ‘Alopeke speckle instrument on the Gemini North 8~m telescope\footnote {https://www.gemini.edu/sciops/instruments/alopeke-zorro/}. ‘Alopeke provides simultaneous speckle imaging in two bands (562 and 832~nm) with output data products including a reconstructed image with robust contrast limits on companion detections (e.g., \citet{Howell_2016}). Five sets of 1000 $\times$ 0.06s exposures were collected and subjected to Fourier analysis in the standard reduction pipeline (see \cite{Howell_2011}). If a nearby companion star is present, interferometric fringes are detected in the Fourier analysis and used to determine the companion separation, position angle, and
magnitude contrast ($\Delta$\textit{m}). Figure~\ref{fig:Gemini_speckle_curve} shows our final 5$\sigma$ contrast curves and reconstructed speckle images. The mean value of each reconstructed image is determined and then radial annuli are made in which each pixel is determined to be above or below the mean. The accumulation of such radial values are used to fit a polynomial curve at a 5$\sigma$ contrast level resulting in the two curves shown in Figure~\ref{fig:Gemini_speckle_curve}. We don't find any companion brighter than $\Delta$m $\sim$ 5 within 1.17\arcsec of the target. At the distance of TOI-1789 (d=223 pc) these angular limits correspond to spatial limits of upto 268 AU.

\subsection{Spectroscopy}
The mass of the planet is determined by precise RV measurements made from the spectrographs PARAS and TCES configured at relatively small aperture-sized telescopes (1-2~m) located respectively at Mount Abu, Physical Research Laboratory (PRL), India,  and Thüringer Landessternwarte Tautenburg (TLS), Germany. We describe these observations in the follwing sub-sections.
\subsubsection{PARAS-PRL}
We acquired a total of 16 spectra of TOI-1789 with the PARAS spectrograph  \citep{Chakraborty_2014} attached to the PRL 1.2~m telescope at the Gurushikhar observatory, Mt. Abu, India. PARAS is a high-resolution (R ${\sim}$ 67000), fiber-fed, pressure and temperature stabilized echelle spectrograph, working in the wavelength range of 380 to 690 nm for precise RV measurements. These spectra were obtained between December 19, 2020 and March 19, 2021 using the simultaneous wavelength calibration mode either with Thorium-Argon (ThAr) or with Uranium-Argon (UAr) hollow cathode lamp (HCL) as described in \cite{Chakraborty_2014} and \citet{uar}, respectively. Only one among the 16 stellar spectra from PARAS was acquired with ThAr HCL, and that was observed on December 19, 2021. After that due to unavailability of pure-Th HCL, we switched to UAr HCL, and the remaining data were acquired with the UAr HCL.

The exposure time for each observation was kept at 1800s which resulted in a SNR per pixel of 12 to 20 at the blaze wavelength of 550 nm. The data reduction were carried out by an automated custom-designed pipeline written in {\tt IDL}, that is based on the algorithms of \citet{2002A&A...385.1095P}. The extracted spectra through this reduction procedure is further used for RV measurements. The RVs were determined by cross-correlating the stellar spectra with a numerical stellar template mask of the same spectral type \citep{1996A&AS..119..373B}. For more details about PARAS data reduction, analysis pipeline, and RV precision stability of the instrument see \citet{Chakraborty_2014, parasfirstplanet}. The offset between absolute RVs derived using ThAr and UAr spectra is found to be 10 m s$^{-1}$, calculated using a standard RV star HD55575 \citep{uar}, and that has been corrected in TOI-1789 RVs for further analysis. All the RVs from PARAS are listed in Table~\ref{tab:rv_table} along with its respective errors. The errors reported here are based on the fitting errors of cross correlation function (CCF) and the photon noise, calculated in the same way as described in \cite{10.1093/mnras/stw1560,Chaturvedi_2018}. 

\subsubsection{TCES-TLS}
A total of 21 useful spectra were obtained with the Tautenburg coude echelle spectrograph (TCES) installed at the 2~m Alfred Jensh telescope at the Thüringer Landessternwarte Tautenburg, Germany between 22 February 2021 and 05 April 2021. TCES is a slit spectrograph with a resolving power of R = 67000 and a wavelength coverage of 470–740~nm. The spectrograph is installed in a Coude room with a temperature
stabilized environment. For details of the observations, one can refer to \cite{2009A&A...507.1659G}. The iodine cell, inserted in the optical path of the spectrograph, acts as a wavelength reference for the spectra. The exposure time for all the obtained spectra were 1800~s each leading to an average SNR per pixel of $\sim$  43 at 564 nm. The standard IRAF routines were used for data reduction that subtract the bias, flat-field the spectra, remove the scattered light, and extract the spectra.
A first-order wavelength calibration is applied through Thorium-Argon calibration lamp which is taken in the beginning and end of each night. Instantaneous wavelength solution is determined by the super-imposed iodine lines. The iodine absorption cell technique requires a complex forward modelling procedure to estimate the Doppler shift in the stellar absorption lines. This technique is instrument specific and depends on the PSF of the spectrograph \citep{1995PASP..107..966V}. RVs are computed by a Python-based software Velocity and Instrument Profile EstimatoR {(VIPER)\footnote{https://github.com/mzechmeister/viper} \citep{Zechmeister2021}}, which is currently being developed as an open-source and a common approach for instrument profile and RV estimation using the Iodine technique. It is based on the standard procedure as described in \citep{1996PASP..108..500B,2000A&A...362..585E}. 

\section{Analysis and Results}

To derive the stellar and planetary properties of TOI-1789 system, we analyzed our photometric and spectroscopic observations in the following subsections. Here we give a brief summary of the procedure and present our results.
\label{sec:analysis}
\subsection{The Host Star}
\subsubsection{Spectroscopic Parameters}\label{sec:sme}

\begin{table}
\centering

\caption{Spectroscopic Properties derived for TOI-1789.}
\label{tab:spec-param}

\begin{tabular}{lccc}

\hline
\noalign{\smallskip}
		Spectroscopic Parameters & $T_{\rm eff}$ & [Fe/H] &  $\log{g_*}$   \\
		 & (K) &  (dex)& (dex)\\
		\hline
		\noalign{\smallskip}

		SME & $5894\pm142$ &  $0.38\pm0.1$ & $4.2\pm0.2$ \\
		\noalign{\smallskip}
		\tt{SpecMatch-Emp} & $5804\pm110$ &  $0.29\pm0.09$ & -- \\
		\noalign{\smallskip}
		\textbf{Global Modelling} & $5984^{+55}_{-57}$ & $0.370^{+0.073}_{-0.089}$ & $3.939^{+0.024}_{-0.046}$ \\
		\noalign{\smallskip}
		\hline
        
\end{tabular}
\end{table}

We determine the spectroscopic parameters by using a relatively high SNR ($\sim$ 65 per pixel) spectrum taken without the iodine cell at TCES-TLS. 
As a first step, we applied the empirical software {\tt {SpecMatch-Emp}} code \citep{2018AJ....155..127H} that compares well-characterised FGKM stars observed with Keck/HIRES to the data, in our case the TCES-TLS spectra with $R = 67000$. We obtain a stellar effective temperature, $T_\mathrm{eff} = 5804 \pm 110$~K, a stellar radius, $R_*$ = $2.078 \pm 0.180$~$R_\odot$, and the abundance of the key species iron relative to hydrogen, $\feh= 0.29 \pm 0.09$ (dex) as mentioned in Table~\ref{tab:spec-param}. To perform a more detailed model, we used  the  spectral analysis package \href{http://www.stsci.edu/~valenti/sme.html}{{\tt{SME}}} \citep[Spectroscopy Made Easy;][]{vp96, pv2017} version 5.22. This code is based on grids of recalculated stellar atmospheric models that calculates a synthetic spectrum. The best fitted stellar parameters are derived through a $\chi^2-$minimisation iteration that compares the synthetic and observed spectra for a given set of parameters. The spectra was synthesised based on the atomic and molecular line data from \href{http://vald.astro.uu.se}{VALD} \citep{Ryabchikova2015} and the Atlas12 \citep{Kurucz2013} atmosphere grids. We refer to \citet{2017A&A...604A..16F} and \citet{2018A&A...618A..33P} for a more detailed description of the modelling. Briefly we chose to model spectral features sensitive to different photospheric parameters: $T_\mathrm{eff}$ from the H$_\alpha$ line wings, and the surface gravity, $\log{g_*}$, from the \ion{Ca}{I} $\lambda \lambda$6102, 6122, and 6162 triplet, and the $\lambda$6439 line. The  abundances of iron and calcium, and the projected stellar rotational velocity, v$\sin{i}$ were fitted from narrow, unblended lines between 6100 and 6500~\AA. We derive the v$\sin{i}$ as $7.0\pm0.5$ km~s$^{-1}$. We fixed the micro- and macro turbulent velocities , $V_\mathrm{mic}$ and $V_\mathrm{mac}$, to 1 and 3~km~s$^{-1}$ \citep{bruntt10,Doyle2014}, respectively. The results from {SME} are in good agreement within the uncertainties with {\tt {SpecMatch-Emp}} (Table~\ref{tab:spec-param}). The SME derived stellar parameters are supplied as priors in the global modelling. The finally adopted stellar parameters obtained through global modelling of the RV and transit data are indicated in the last row of Table~\ref{tab:spec-param} in bold font. For more details on the global modelling, see Section~\ref{sec:exofast}. One thing to note here is that we find the TOI-1789 is a late F-type star with our spectral analysis. Nevertheless, SIMBAD reported it as `K0' spectral type.

\subsubsection{Rotational Period Determination}
\label{subsubsec:Prot}
\begin{figure}
	\includegraphics[width=\columnwidth]{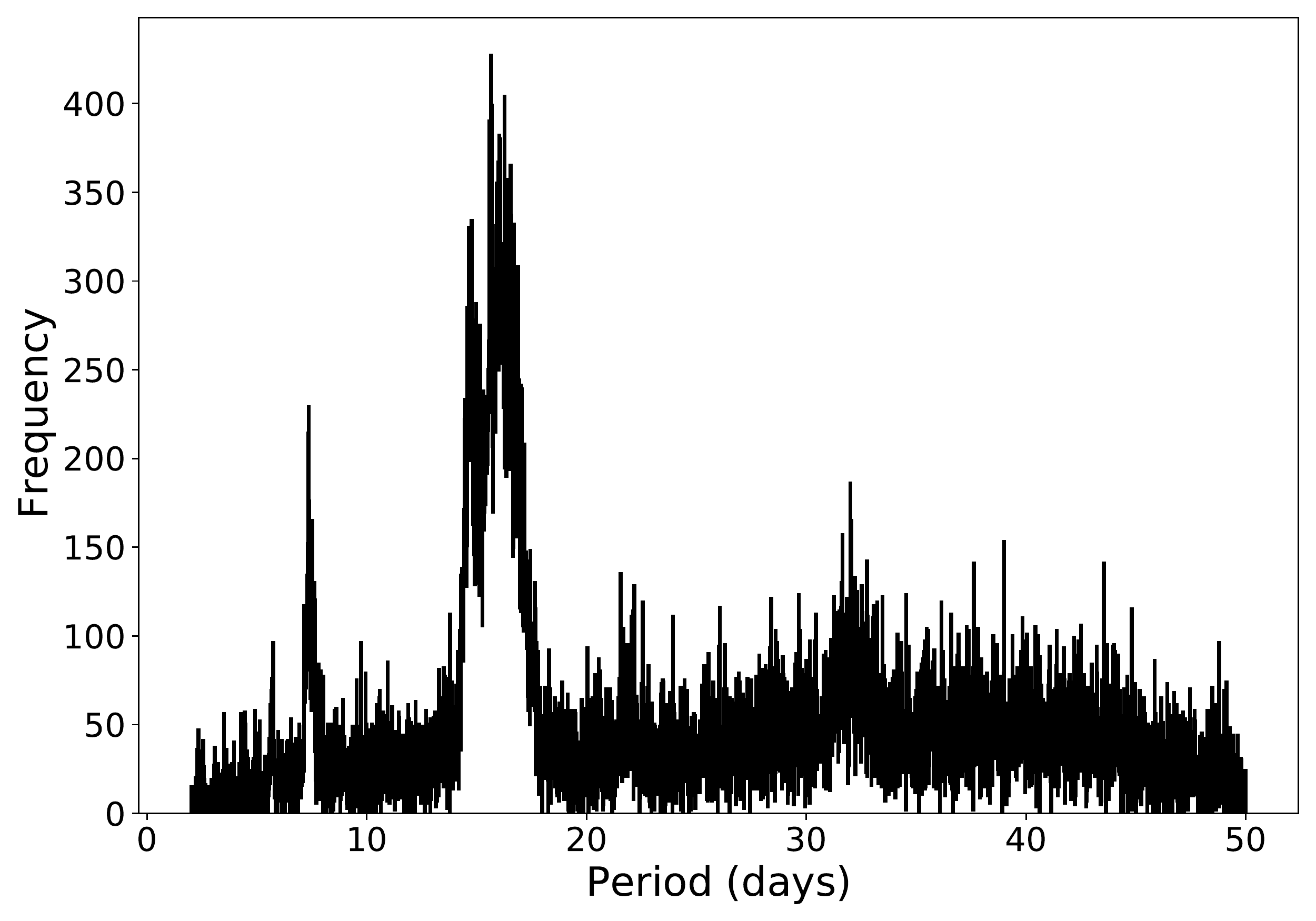}
    \caption{Histogram for the posterior distribution function(PDF) for the estimated rotational period of the star (Period) while using the combined and nightly binned photometric datasets from ASAS, SuperWASP, KELT. We see the PDF peaking at 16 days with other less significant peaks at 8 and 32 days.}
     \label{fig:GP-Prot}
\end{figure}
TOI-1789 is a moderately rotating star as determined through SME spectral modelling for v$\sin{i}$ of $7.0\pm0.5$ km~s$^{-1}$. \textit{}
We use the radius of $2.168^{+0.036}_{-0.034}$ \(R_\odot\) (from global modelling Section.~\ref{sec:global}) to calculate the rotational period of about 15.7 days by assuming that the star is observed equator-on (\textit{i} = 90$^{\circ}$). Due to the presence of active regions or spots on the surface of the star, the rotation of the star often produces a non-sinusoidal signal. The lifetime of these signals is not well constrained and the differential rotation of the star makes this signal quasi-periodic \citep{2011A&A...525A.140D, 2014MNRAS.443.2517H}. These activity-induced signals are also introduced in the RV data, making it a nuisance for the planet detection. Such activity signals often cannot be modelled through a plain sinusoidal function. The non-parametric Gaussian processes can be used to model these signals \citep{2018MNRAS.474.2094A}. 

We retrieved the publicly available SuperWASP\footnote{\url{https://wasp.cerit-sc.cz/form}}, ASAS3\footnote{\url{http://www.astrouw.edu.pl}} and KELT\footnote{\url{https://exoplanetarchive.ipac.caltech.edu/docs/KELT.html}} photometry data for this source. The ASAS3 data were observed between Dec 2002-May 2009, the SuperWASP data were observed between May 2004-October 2007, and the KELT lightcurves were observed between October 2006-November 2008. We nightly binned each dataset separately and flagged them based on their instruments before supplying these as an input to model the GP. We invoked the quasi periodic \texttt{george} kernel \citep{Ambikasaran2015} which is built-in the \texttt{juliet} routine \citep{juliet} and applied it on this combined dataset.  The \texttt{george}  kernel uses the exponential sine-squared  function multiplied with a squared-exponential function of the form
\begin{equation}
\centering
~~~~~~~~~~~~~~~~k(\tau) = \sigma^{2}_{GP}~exp (-\alpha_{GP}~\tau^2 -\Gamma sin^2{ (\frac{\pi \tau}{P_{rot}})})
\end{equation}

where $\sigma_{GP}$ is the GP amplitude in parts per million (ppm), $\Gamma$ is the dimensionless amplitude of the GP sine-squared component, $\alpha$ is the inverse length scale of the decay times (day$^{-2}$) component, $\tau$ is the time lag, and P$_{\rm {rot}}$ is the rotational period of the star in days. We used separate and uninformative priors for the photometry datasets from different instruments with a broad range of prior parameters as discussed in \cite{2020A&A...643A.112S}. We kept the rotational period uniformly varying between 2-50 days. We display the histogram plot for the posterior density function (PDF) for the successful run. As seen from  Figure~\ref{fig:GP-Prot}, we see the most significant peak at around 16 days. The amplitude of variation is 7~mmag during peak stellar variability. There are other less significant peaks seen at 8 and 32 days which could be possibly related to the rotational period of the star. Since, our derived P$_{\rm {rot}}$ from the v~$\sin{i}$ is also 15.7 d, we can conclude from this analysis that the P$_{\rm {rot}}$ for TOI-1789 is between 15--16 days.

\subsubsection{Bisector Analysis}\label{sec:bis}
\begin{figure}
	\includegraphics[width=\columnwidth]{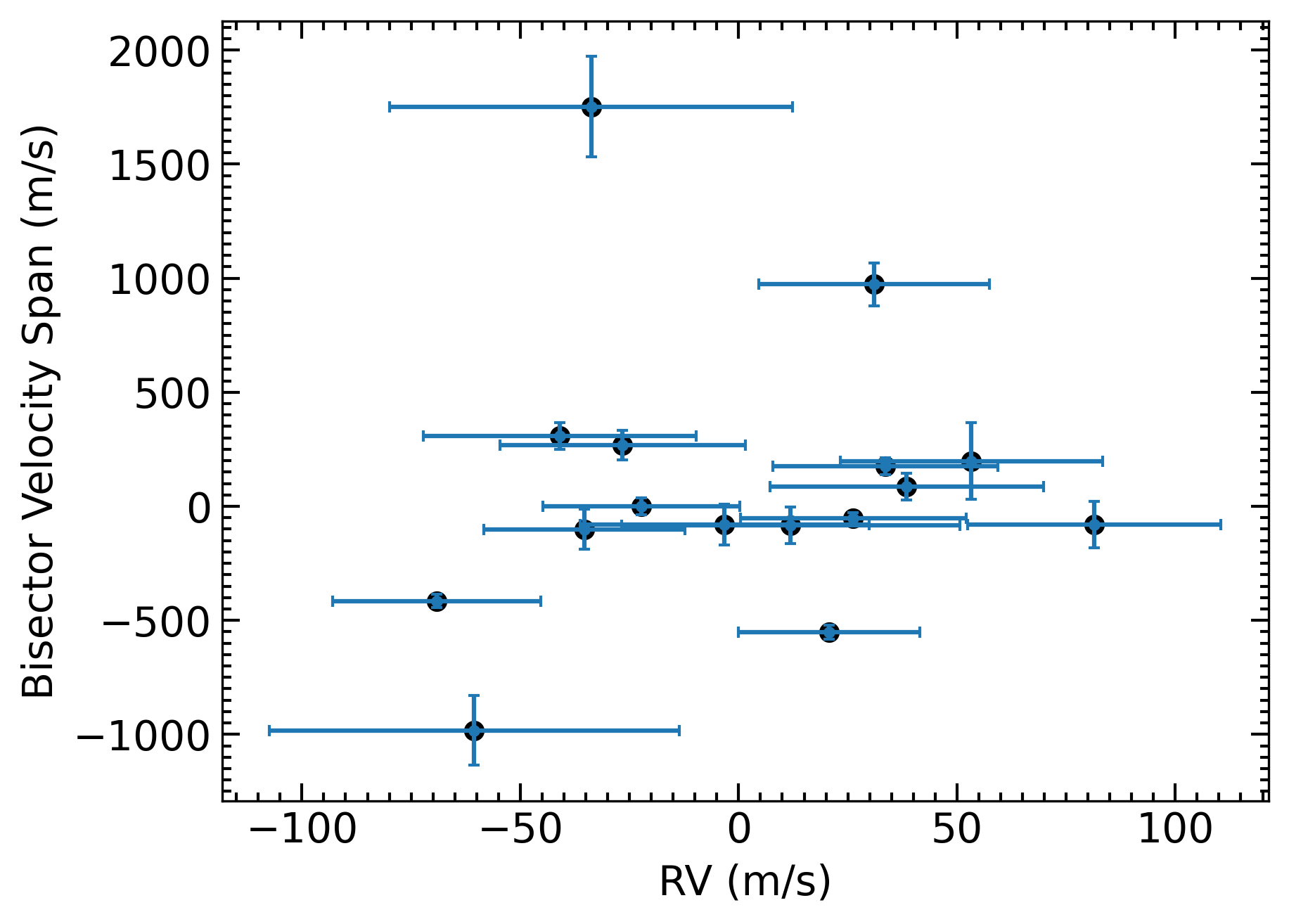}
    \caption{Plot of Bisector analysis from PARAS data. No significant correlation was observed between RV and BIS. (Correlation coefficient $\approx$ 0.15, p-value = 0.58).}
    \label{fig:BIS}
\end{figure}
To investigate the cause of the RV variations in the PARAS data, we did the bisector analysis. The bisectors are calculated from the PARAS CCFs, where the CCF represents the average profile of the lines in the stellar mask used to calculate RVs. As described in \cite{2005A&A...442..775M}, we calculate the bisector of each horizontal segment of the CCF,  which are at the same flux level from left to right. The line bisector is the combination of all these bisectors from the core to the wings of the CCF. Then the bisector velocity span (BIS) is calculated by taking the difference of average velocity values between the top and bottom zones of the CCF (BIS $=$ V$_{top}$ - V$_{bottom}$). This difference (BIS) represents the symmetry of the CCF or the absorption lines in the spectra. A strong correlation, negative or positive, between RVs and BIS corresponds to the RV signal being attributed to stellar activity caused by spots or by other atmospheric phenomena like granulation, pulsation or turbulence, or due to likely contamination in the spectra from a nearby star \citep{2001A&A...379..279Q, 2005A&A...442..775M}.

As described in Section~\ref{sec:obs}, RV data is in phase with the transit ephemeris. From Section~\ref{subsubsec:Prot}, the estimated rotation period is between 15-16 days a value much different than the planet period of 3.21 days. As it can be seen in Figure~\ref{fig:BIS}, we don't find any significant correlation (Correlation coefficient $\approx$ 0.15, p-value = 0.58). Thus activity related correlation seems unlikely. Moreover, we do not see any likely contamination of the target star from high-resolution imaging and neither any signature of blending in the bisector analysis. Thus, we conclude that the RV variations are due to the planetary signature seen in the data.  The values for RVs, BVS and their respective error bars are listed in Table~\ref{tab:rv_table}, and the BVS plotted against RVs are shown in Figure~\ref{fig:BIS}. 

\subsection{Periodogram Analysis}

\begin{figure}
    \centering
	\includegraphics[width=\columnwidth]{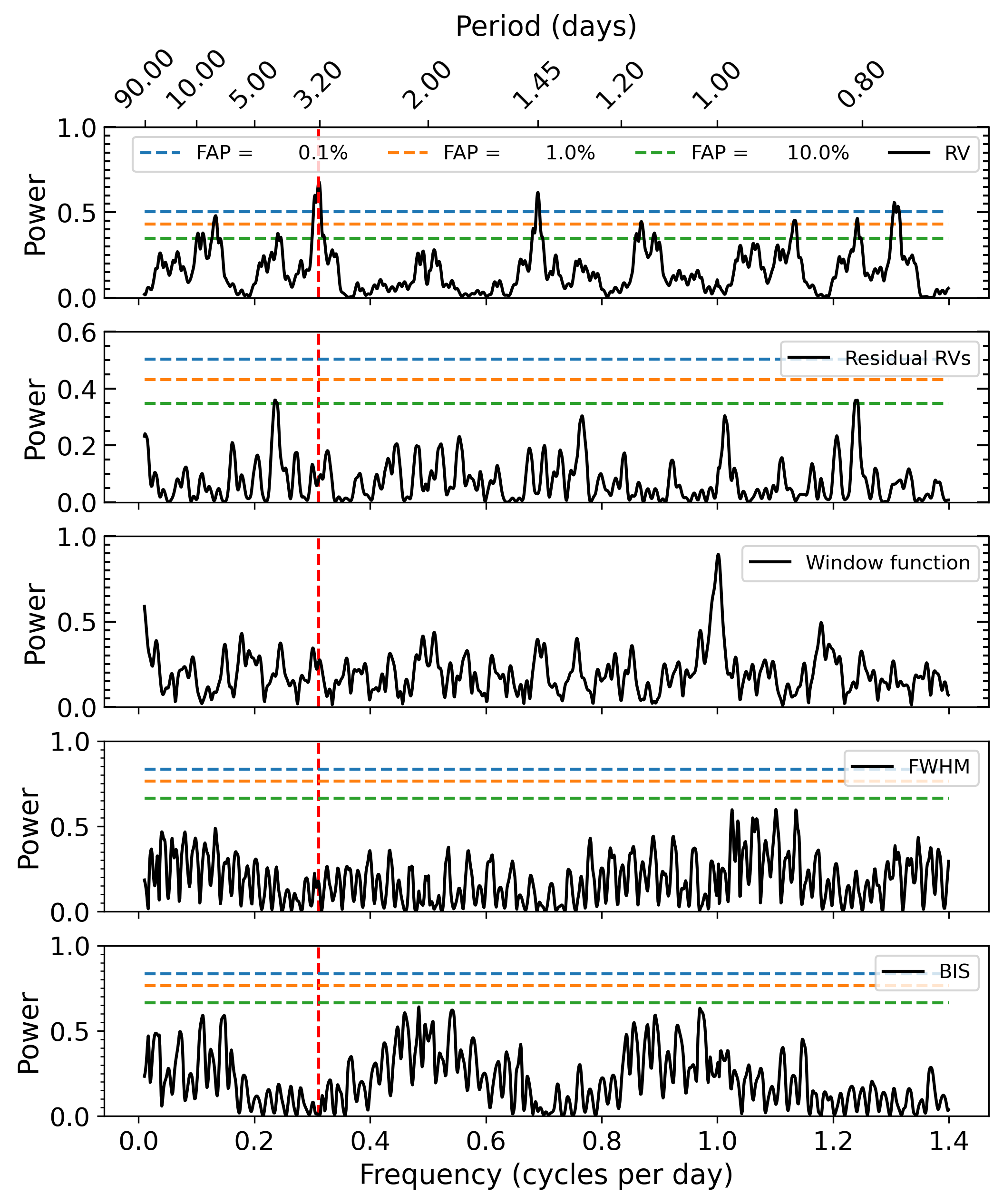}
    \caption {The GLS periodogram for the RVs, residual RVs, window function, FWHM, and bisector span of TOI-1789 is shown in panel \texttt{1}, \texttt{2}, \texttt{3}, \texttt{4}, and \texttt{5} (upper to lower) respectively. The primary peak is seen at a period $\approx 3.21$ days (dashed red line), consistent with the orbital period of the planetary candidate obtained from photometry. The FAP levels (dashed lines) of 0.1$\%$, 1$\%$, and 10$\%$ for the periodograms are shown in the legends in the panel~\texttt{4}. \textbf{Note.} For periodogram analysis of the RVs, the whole dataset was used. In contrast, only PARAS data was used for the last two analyses as superimposed iodine lines are presented in TCES spectra.}
    \label{fig:periodogram}
\end{figure}

We show the computed Generalized Lomb-Scargle periodogram \citep[GLS;][]{periodogram} for RVs, residual RVs, window function, FWHM, and BIS in panel \texttt{1}, \texttt{2}, \texttt{3}, \texttt{4}, and \texttt{5} (top to bottom) respectively in the Figure~\ref{fig:periodogram}. For finding the periodicities in our RV data, we combined RVs from both the spectrographs, PARAS and TCES. Firstly, we corrected the instrumental offsets from RVs by subtracting corresponding average RV values and then computed the periodogram (panel~\texttt{1}). Here, we use the equations given in \citet{periodogram} to normalize the periodogram and calculate false alarm probability (FAP). We consider the threshold FAP of 0.1\% for any significant signal and find the most significant signal at a period of $\sim$ 3.21 days (marked as vertical dashed line in Figure~\ref{fig:periodogram}), the same as the orbital period derived from the transit data (both ground and space-based) for this planetary candidate. We evaluated the FAP at this periodicity using a bootstrap method of 1000000 randomization, over a very narrow range, centering at this period. It gives an FAP of 0.007\%, which strongly confirms a periodic signal in our RV data set. 
There are other two significant signals in the RV periodogram at a period of $\sim$ 0.76 and $\sim$ 1.45 days that vanish when we subtract this $\sim$ 3.21 days periodic signal using a best-fit sinusoidal curve, as can be seen in the residual periodogram (panel~\texttt{2}). These are the 1-day aliases of the orbital frequency (f$_{orb}$) or the 3.21 days signal. Specifically, 1/0.76 is the 1-day alias of f$_{orb}$, and 1/1.45 is the 1-day alias of -f$_{orb}$. The residual periodogram does not show any other significant periodicity with FAPs above our threshold of 0.1\%. The spectral window function is shown in the panel~\texttt{3}.
Furthermore, we also computed the GLS periodogram of the CCF FWHMs and BIS using PARAS data, which is shown in panel~\texttt{4} and panel~\texttt{5}, respectively (for more information of bisector analysis, please see Section~\ref{sec:bis}). These diagnostics are generally used as stellar activity indicators since they quantify the line asymmetries which mimic the Doppler shifts. We do not find any statistical significant signals of the stellar activity in the periodograms. Here, we could not use the TCES data as it has superimposed iodine lines in the spectra.

\subsection{Global Modelling}\label{sec:global}

We used EXOFASTv2 \citep{exofastv2} to constrain the host star parameters in the global model and find the orbital and planetary parameters of the system. The EXOFASTv2 uses the Differential Evolution Markov Chain Monte Carlo (MCMC) technique to fit the multi-planetary systems with multiple RV and photometry datasets and at the same time provides the opportunity for deriving stellar parameters using spectral energy distribution (SED) and isochrones. It also diagnoses the convergence of the chains using built-in Gelman-Rubin statistics \citep{Gelman_rubin1992, Gelman_rubin2006}.

\begin{figure}
	\centering
	\includegraphics[width=7cm, height=7cm, trim={1cm 0 0 0}]{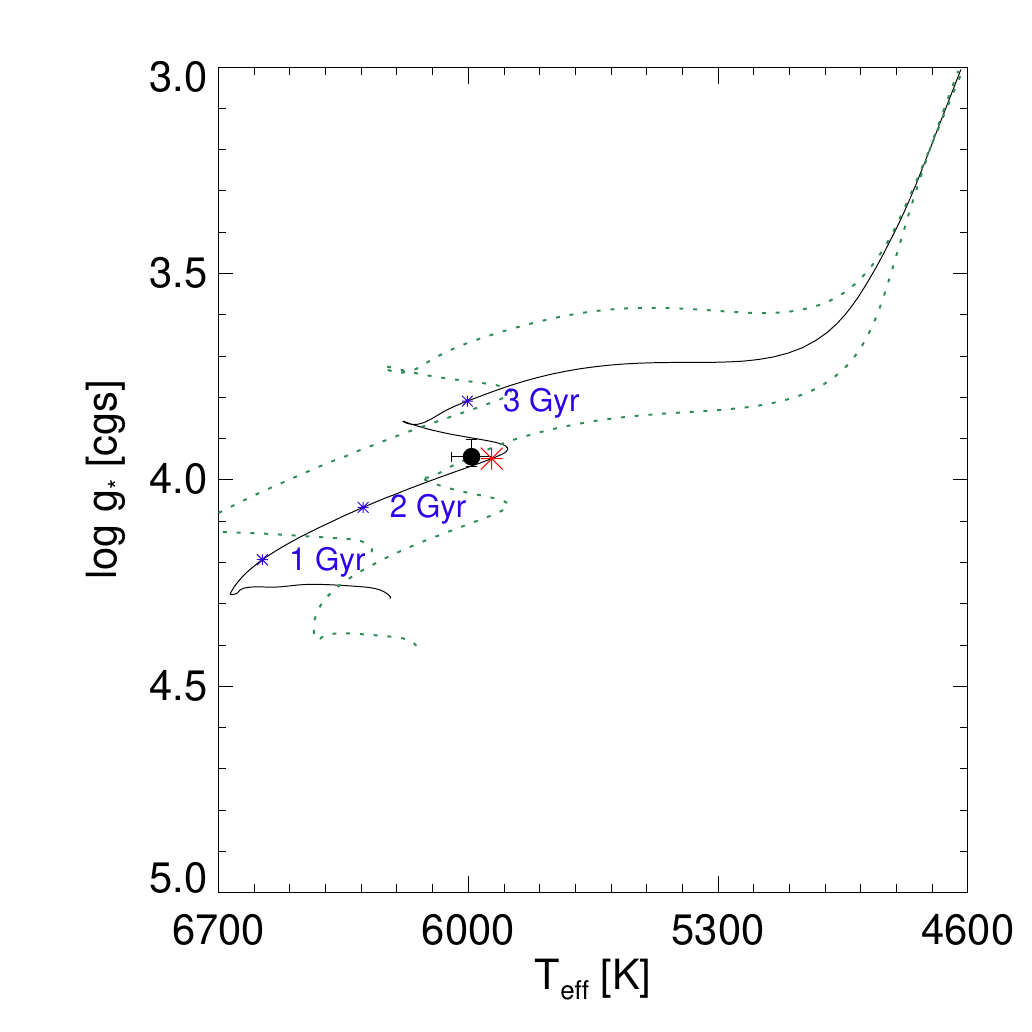}
    \caption{The solid black line depicts the most likely evolutionary track for TOI-1789 from MIST. The black circle is at the model value \teff\space and $\log{g_*}$ with its errorbars. The red asterisk represents the model value for Equal Evolutionary Point (EEP) or the age of TOI-1789 along the track. The other two dashed lines are the evolutionary tracks of masses 1.36 \(M_\odot\) and 1.64 \(M_\odot\), whereas blue asterisks represent the age values 1 Gyr, 2 Gyr and 3 Gyr along the track of TOI-1789.}
    
    \label{fig:mist}
\end{figure}

\subsubsection{Modelling the host star}\label{sec:sed}
\begin{figure}
    \centering
	\includegraphics[width=\columnwidth]{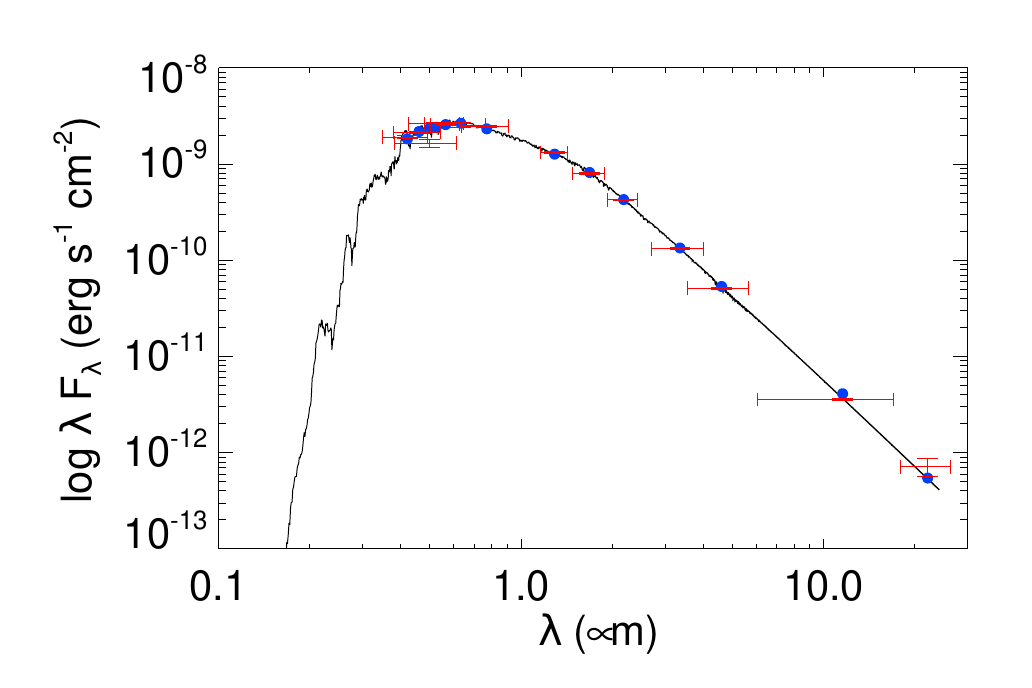}
    \caption{The spectral energy distribution of TOI-1789. The Red markers with the horizontal error bars denote the photometric measurements in each filter and their bandwidth, while the vertical error bars shows the measurement uncertainty.
    The Black curve is representing the best fit Kurucz stellar atmosphere model with highlighted blue circles as the model fluxes over each passband.}
    \label{fig:sed}
\end{figure}

We derive the host star parameters using the MIST isochrones \citep{mist_choi, mist_dotter} and the Spectral Energy Distribution (SED) fitting \citep{SED1} using the Kurucz stellar atmosphere model \citep{Kurucz} within the EXOFASTv2 framework. The SED fitting combined with MIST isochrones and transit precisely determine the mass, radius, $\log{g_*}$, and age of the star \citep{Torres}. Therefore, we use the TESS and ground based transit data to constrain the stellar parameters. We place gaussian priors on $\teff$ and \feh, derived from the spectral analysis, and parallax from $Gaia EDR3$ \citep{gaiaedr3}. A uniform prior with an upper limit on $V$-band extinction from the \citet{extinction} and dust maps at the location of the host star are also applied. We plot the $\log{g_*}$ versus \teff\space corresponding to Table~\ref{result_exofast} in Figure~\ref{fig:mist} and show the most likely MIST evolutionary track for TOI-1789 (solid line). We also show the evolutionary tracks (dashed lines) of two different masses 1.36 \(M_\odot\) and 1.64 \(M_\odot\) ($\pm$ 0.14 of the mass of TOI-1789) in the Figure. We used broadband photometry from Tycho BV \citep{tycho}, APASS DR9 BV, SDSSgri \citep{APASS}, 2MASS JHK \citep{JHK}, ALL-WISE W1, W2, W3, and W4 \citep{ALLWISE}, which are listed in Table~\ref{tab:star_table}. The best fitted SED model along with these broadband photometry fluxes are shown in Figure~\ref{fig:sed}.

The PDF of stellar mass and age (and their correlated parameters) show bimodality,  which can be seen in Figure~\ref{fig:corner}. This type of bimodality has been observed in several recent studies \citep{ngts13,bimod2,bimod3,bimod4,bimod5} and it appears due to the degeneracy between the MIST isochrones in the region of \teff -- $\log{g_*}$ plane occupied by TOI-1789. The two peaks in the PDF are centered at a mass of 1.35 \(M_\odot\) (age = 4.15 Gyr) and 1.51 \(M_\odot\) (age = 2.71 Gyr) with the probability of 30$\%$ and 70$\%$, respectively. We finally adopted the solution provided by EXOFASTv2, which is centered at the most probable values for M$_{*}$, age, and their correlated parameters and thus have larger uncertainties considering the bimodality. The adopted stellar parameters along with 1$\sigma$ uncertainties are displayed in Table~\ref{tab:spec-param}. 

The final adopted values for the host star   are $\log{g_*}$ = $3.943^{+0.023}_{-0.043}$, {\ensuremath{\,M_*}} = $1.507^{+0.059}_{-0.14}$ \(M_\odot\), {\ensuremath{\,R_*}} = $2.168^{+0.036}_{-0.034}$ \(R_\odot\), at an age of $2.73^{+1.3}_{-0.51}$ Gyr. The $\log{g_*}$ determined here is slightly lower ($\sim$ 1.3$\sigma$) than the one estimated with SME (Section~\ref{sec:sme}). Stellar gravity is more accurately determined with SED, isochrones combined with the transit data which places a strong constraint on the stellar density of the star \citep{2017AJ....153..178S}. Therefore we adopt the stellar parameters obtained from this global modelling to determine the orbital and planetary parameters. 

\subsubsection{Orbital Parameters}\label{sec:exofast}
The independent fit of one data set (either RV or transit) is useful in constraining mutually independent parameters obtained from either of these datasets. For e.g., $b$, $i$, $R_{p}$, and $a$ are exclusively dependent on the transit data, while $K$ and hence $M_p$ depends on the RV data. However, the parameters like $P$, $T_{c}$, $\omega$, and $e$ depend on both the datasets and can be well constrained by fitting them simultaneously.

\begin{figure}
\centering
\begin{subfigure}{\columnwidth}
  \includegraphics[width=\columnwidth]{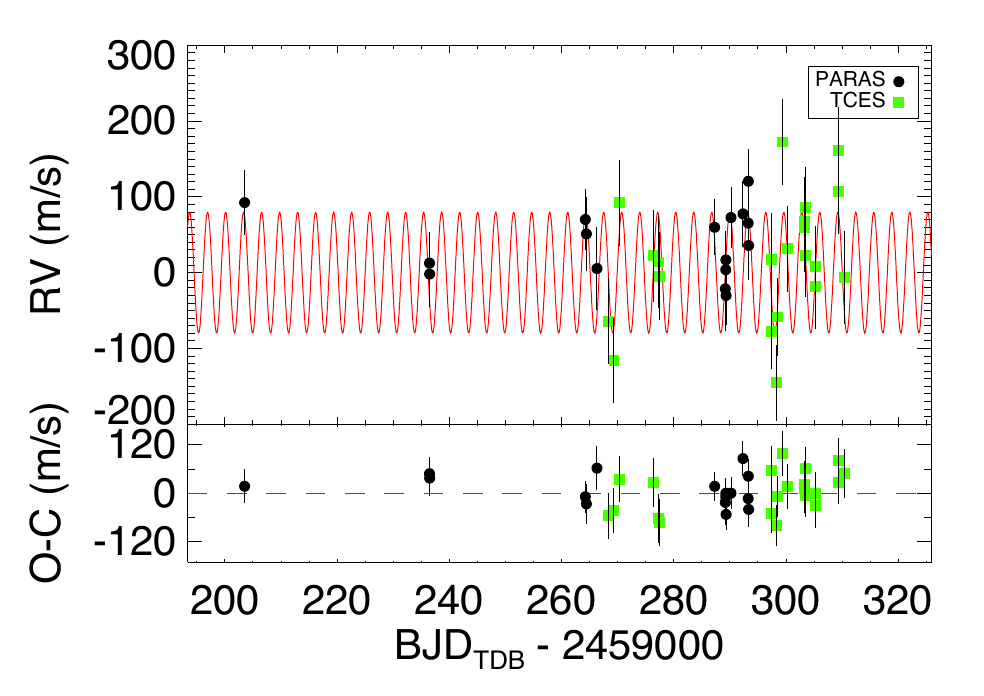}
\end{subfigure}

\begin{subfigure}{\columnwidth}
  \includegraphics[width=\columnwidth]{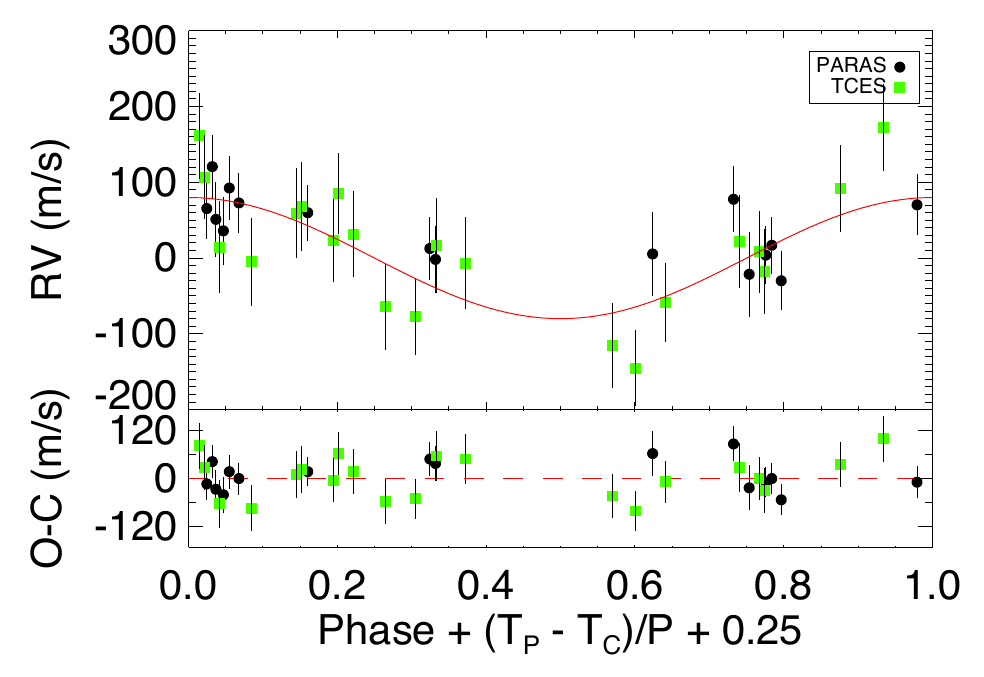}
\end{subfigure}
\caption{TOI-1789 RVs from PARAS (black dots), and TCES (green squares) are plotted here with respect to time in the upper panel, and with respect to orbital phase ($\sim$ 3.21~d) in the lower panel, after subtracting the corresponding instrumental RV offsets $\gamma_{rel}$, listed in Table~\ref{result_exofast}. The red line represents the best-fit RV model for the planet from EXOFASTv2 (see Section~\ref{sec:exofast}). The bottom-panel shows the residuals between the data and the best-fit model.}
    \label{fig:RV_curve}
\end{figure}

We use the photometry data from TESS and PRL (two datasets, Section~\ref{sec:prl_photo}), and RV data from PARAS and TCES for the simultaneous fitting. We provide the Gaussian priors on the stellar parameters derived in Section~\ref{sec:sed}, along with the starting values for $P$ and $T_{c}$ given by the TESS QLP pipeline. We also include offsets and jitter terms for each dataset of each instrument in RV and LC fitting.
The transit model is generated using the \citet{Mandel2002}; re-sampling the transit model over 10 steps to account for the TESS long cadence (30 min) data \citep{kipping}. The photometric data were modeled with a quadratic limb-darkening (LD) law for $TESS$ and Bessel-$R$ (PRL transits) passbands. The $u_{1}$ and $u_{2}$ are derived by the interpolation of the \citet{Claret,Claret_tess} LD models. We used 56 chains and 50000 steps for MCMC global fitting, which allowed the fit to converge. First, we fit a circular orbit model keeping eccentricity to zero, and then an eccentric orbit model, by keeping the \ecosw\space and \esinw\space  as free parameters to check for any significant eccentricity in the orbit. The eccentricity is found to be $0.1\pm0.075$. We calculated the Akaike Information Criterion (AIC, \citet{AIC}) as well as Bayesian Information Criterion (BIC, \citet{BIC}) for both the models and find the $\Delta$AIC between the models to be 6.0, moderately favouring the circular model over the eccentric model, while the $\Delta$BIC to be 17.0, strongly supporting the circular orbit model. Thus, we adopted the circular orbit model and report the orbital and planetary parameters in Table~\ref{result_exofast}. We find that TOI-1789~b has a mass of $0.70\pm0.16$ \mj~and a radius of $1.44^{+0.24}_{-0.14}$ \rj, which corresponds to the density of $0.28^{+0.14}_{-0.12}$ g cm$^{-3}$. The resulting best fit models from EXOFASTv2 for the transit light curves are plotted in Figure~\ref{fig:TESS_light_curve}, and for the RVs in Figure~\ref{fig:RV_curve}. Figure~\ref{fig:corner} shows the covariances for all the fitted parameters for the global joint fit. For TOI-1789, b+R$_{p}$/R$_{*}$ > 1 represents a grazing transit configuration. In such cases, the posterior probability densities contain a strong degeneracy between the transit impact parameter b (and its related parameters such as inclination, etc.) and the planetary radius R$_{p}$/R$_{*}$. Due to this the planetary radius can not be determined precisely. It can be seen for TOI-1789 in Figure~\ref{fig:corner}, there is degeneracy between R$_{p}$/R$_{*}$ and inclination $\cos{i}$. Although we have used short cadence data along with TESS data and the transit is deeper (0.25$\%$), the radius is not well-constrained, having 9-16$\%$ uncertainty.

\section{Discussion}
\label{sec:discussion}
TOI-1789, with its precise determination of host star and planet properties (with the precision of 22$\%$ and 9-16$\%$ in planetary mass and radius, respectively), makes an important contribution to the study of gas giant planets around slightly evolved stars. In our discussion, we focus on the evolutionary status of the host star and the derived planet properties.

\subsection{The evolved star}\label{sec:evolved}
Semi-analytical disk models indicate that the frequency of giant planets must increase with the mass of the host star between 0.2--1.5 \msun$_\odot$ \citep{2008ApJ...673..502K, 2005ApJ...626.1045I}. However, this trend is expected to decrease above 1.5 \msun$_\odot$ due to a smaller growth rate, longer migration timescale, and shorter lifetime of the protostellar disk \citep{2015A&A...574A.116R}. Looking for planets around main sequence stars more massive than the sun can help shed some light on this aspect. These stars have few spectral lines for Doppler measurements and are often broadened by the rapid rotation of the star. This has been the reason for RV surveys to have traditionally targeted slow rotating FGK type stars. However, as rapidly rotating stars evolve off the main sequence they slow down considerably and become much cooler making it relatively easy to search for planets around them. This fact was exploited by dedicated planet searches around intermediate mass sub-giants leading to dozens of planet discoveries \citep{2007ApJ...665..785J, 2010PASP..122..701J, 2011ApJS..197...26J}. An important result obtained from the survey of giant stars at the Lick observatory pointed out that the occurrence rate peaks at a stellar mass of $1.9^{+0.1}_{-0.5}$ \msun$_\odot$. However, many of the discovered planets around evolved stars were found at large orbital separations \citep{2003ApJ...599.1383H, 2007ApJ...669.1336F, 2007ApJ...670.1391R, 2008ApJ...675..784J}. This is not a surprise as star-planet interaction is largely governed by tidal forces. When the stellar rotational period is longer than the planet orbital period, the star experiences spinning up, leading to orbital decay. Synchronization and circularization of orbit occurs in systems where the total angular momentum exceeds a critical value. When this total angular momentum is small enough, the orbit of the planet can continue to shrink and be engulfed by the host star. This phenomenon entirely depends on the dissipation time scales for the star (\cite{2008EAS....29....1M} and references therein). The role of tidal forces becomes increasingly important in the context of host stars being in an evolved state. There is a higher chance of the planet being destroyed by the evolved star \citep{2011ApJ...737...66K, 2013ApJ...772..143S}. However, there is no obvious way to estimate these tidal dissipation forces. The circularization timescale for such planets can be used to quantify tidal dissipation inside planets \citep{2010ApJ...723..285H, 2012ApJ...750..106S}. Most of the discovered hot Jupiters with periods less than 3 days are found to be on circular orbits. We calculate the circularization timescales for the orbit of TOI-1789~b, which is $\tau _{cir} $ = 0.08 Gyr  (for $Q_P = 10^6$, equation (3) of \cite{circ_time}). This is less than the age of the star as calculated from our work (Section~\ref{sec:sed}). 

\begin{figure}
	\includegraphics[width=\columnwidth]{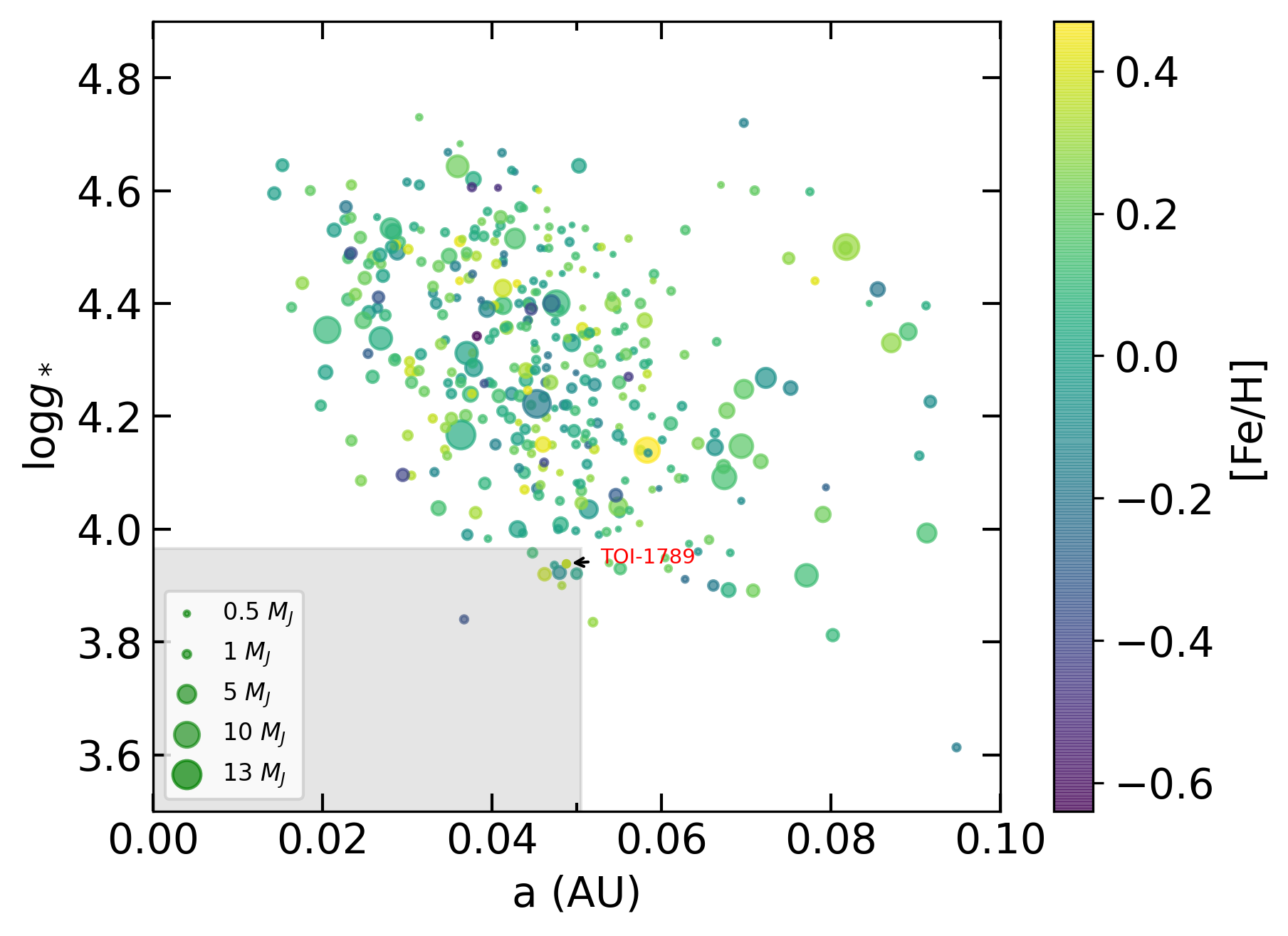}
    \caption{The surface gravity ($\log{g_*}$) of host stars having the transiting hot Jupiters ($0.25M_{J}<M<13M_J$ and P<10 days) in their orbits are plotted as a function of orbital separation ($a$) in AU. Datasets used in this diagram are taken from the TEPcat database$^{\ref{tepcat}}$, and only planets with uncertainty on the masses and radii <25$\%$ are considered. Metallicity of the host star is colour-coded, with colours ranging from low metallicity (in blue) to high metallicity (in yellow). The size of individual dots represent the planet's mass in Jupiter mass. The grey shaded region represents the boundary (a $\leq$ 0.05 AU $\&$ $\log{g_*}$ $\leq$ 4.1 dex) and the other 7 exoplanets as described in Section~\ref{sec:evolved} can be seen under this shaded region. The position of TOI-1789 marked with arrow and is labeled as red text in the figure.}
    \label{fig:rel}
\end{figure}

Figure~\ref{fig:rel} shows all the transiting hot Jupiters taken from the Transiting ExoPlanet catalogue (TEPcat) database\footnote{\label{tepcat}https://www.astro.keele.ac.uk/jkt/tepcat/ \cite{10.1111/j.1365-2966.2011.19399.x} as of June 9, 2021} with uncertainties on the masses and radii < 25\%. We use a planet mass cut off between $0.25M_{J}<M<13M_J$ and orbital period less than 10 days to stay consistent with the definition of hot Jupiter \citep{lower_mss_limit, upper_mass_limit}. We plot $\log{g_*}$ as an indicator of the evolutionary status of the star versus the separation of the host star ($a$) from its orbiting planet in AU. Stars are colour coded according to their metallicity. Stars with high metallicity appear as yellow, whereas the ones with sub-solar metallicities are blue. The size of the individual points represent the planet's mass. We mark the position of TOI-1789 with the arrow, and is labeled as red text in the figure. One notes that close-in transiting planets are more common around the main sequence stars ($\log{g_*}\ge~4.1$). Another thing to notice is that most of the planet host stars have super-solar metallicity. This correlation of stellar metallicity with planet occurrence rate for main sequence stars was reported early on by \cite{1997MNRAS.285..403G, 2004A&A...415.1153S, 2005ApJ...622.1102F}. There were surveys like the Next 200 Stars (N2K) and ELODIE survey which were particularly designed to utilize this information discovering several Jovian exoplanets \citep{2005ApJ...620..481F, 2005A&A...444L..15B, 2006A&A...458..327M}. The study of the planet-metallicity relation was extended to evolved stars thanks to the survey of giant stars carried out at the Lick Observatory \citep{2001PASP..113..173F}. A strong planet-metallicity correlation with a power-law exponent of $1.7^{+0.3}_{-0.4}$ was reported from this survey \citep{2015A&A...574A.116R}. With a derived metallicity of $0.373^{+0.071}_{-0.086}$ dex, TOI-1789 joins the many other metal-rich host stars.

Given any stellar type, there is a detection bias favouring close-in planets. We see from Figure~\ref{fig:rel} that there are fewer planets situated to the proximity of their host stars for the case of evolved stars (logg<4.1) as compared to the main-sequence stars. This dearth of close-in planets around sub-giants and giant stars was also noted by \cite{2010ApJ...709..396B}. The surface gravity of host star ($\log{g_*}$) shows a weak linear correlation with $a$ as seen from this figure (Pearson correlation coefficient = -0.34, $p = 8.2\times10^{-12}$). This plot is representative of the fact that the proximity of the orbiting planets depends on the evolutionary status of the host star. TOI-1789~b is fairly close-in located at a distance of 0.048 AU from its host star and seems to lie very close to this boundary. This boundary could be dictated by the fact that such close-in planets could be engulfed by its host star in the course of evolution on its sub-giant/giant phase. This fact is also supported by the masses of the planets seen close to this boundary limit. Most of these close-in planets have sub-Jovian to Jovian-like masses. Tidal forces are expected to be larger when the orbiting planet has a larger mass \citep{2008EAS....29....1M}. As of now, only 8 planetary systems (including this work) around stars similar to or more evolved than TOI-1789 are known, which are orbiting closer to their host star (a $\leq$ 0.05 AU). These are namely, (HATS-12: \cite{hats12}, HATS-26: \cite{hats26}, HATS-40: \cite{hats40}, WASP-71: \cite{wasp71}, WASP-78: \cite{wasp78}, WASP-82: \cite{wasp82} and WASP-165: \cite{wasp165}).
We mark this boundary (a $\leq$ 0.05 AU $\&$ $\log{g_*}$ $\leq$ 4.1 dex) with the shaded grey region in Figure~\ref{fig:rel}.
At the age of $\sim$ 2.7 Gyr, TOI-1789 is a slightly evolved late F-type star with  $\log{g_*}$ of $\sim$ 3.9, a radius of $\sim$ 2.2 \(R_\odot\) and a mass of $\sim$ 1.5 \msun$_\odot$. 
The circular orbit for the planet seen here is evidence of the strong tidal interaction due to its host star. This makes TOI-1789~b a rare and important system in understanding the evolution of close-in planets around stars transiting off their main sequence. 

\subsection{The heated planet}

\begin{figure}
	\includegraphics[width=\columnwidth]{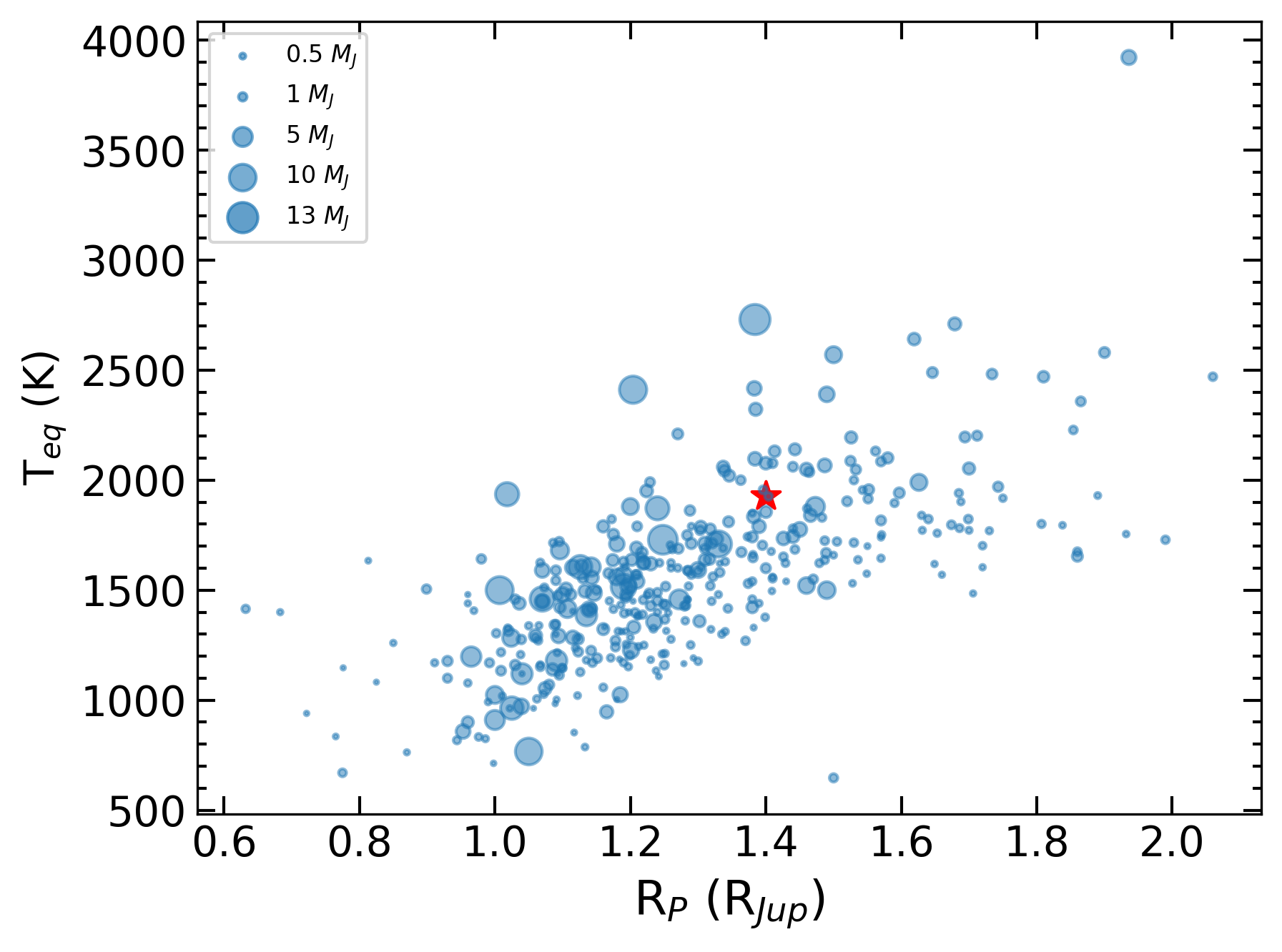}
    \caption{The Planetary Radius-Equilibrium temperature relation diagram for known transiting hot Jupiters ($0.25M_{J}<M<13M_J$ and P<10 days).  Data are taken from the TEPcat database$^{\ref{tepcat}}$, only planets with uncertainty on the masses and radii <25$\%$ are considered. The size of individual dots correspond to the mass of the planet in terms of Jupiter. The position of TOI-1789~b is plotted as red asterisk symbol.}
    \label{fig:relation}
\end{figure}

Understanding the radius inflation of transiting hot giants is an important aspect of exoplanetary science. The determination of precise stellar ages is useful in this context. Finding inflated Jupiters during the end-of-main sequence phase is indicative of a deep heating mechanism. However, if hot Jupiters
are found to be inflated during the post-main-sequence, then heating must be concentrated at deposition pressure $\ge 10^{5}$ bars \citep{2020ApJ...893...36K}. Considering the aspects of both stellar irradiation and the age of the system, \citet{2020ApJ...893...36K} postulated that inflated Jupiters would be the most common around stars with masses 1.0~\msun$_\odot$<$M_*$<1.5~\msun$_\odot$, which is also true for TOI-1789 which has a mass of $1.507^{+0.059}_{-0.14}$ \msun$_\odot$ and the radius of the planet of $1.44^{+0.24}_{-0.14}$ \(R_J\). Figure~\ref{fig:relation} shows the data for the known transiting hot Jupiters from the TEPcat database,$^{\ref{tepcat}}$ where we plot the planet radius against its equilibrium temperature. The size of the individual dots on this figure corresponds to the mass of the planet. We mark the position of TOI-1789~b as a red asterisk symbol. It has a fairly inflated radius relative to its mass. The equilibrium temperature for the planet (assuming no albedo and perfect redistribution) is estimated to be $1927^{+27}_{-17}$ K.
Several studies indicate that inflation in hot Jupiters would be due to the high incident fluxes received by them through their host stars \citep{2002A&A...385..156G, 2017ApJ...841...30T, 2019A&A...632A.114S}. The total incident flux received by TOI-1789~b is $3.13^{+0.18}_{-0.11} \times 10^9$ erg s$^{-1}$ cm$^{-2}$. This flux-limit is more than 15 times the threshold (2 $\times 10^8$ erg s$^{-1}$ cm$^{-2}$) as described in \cite{2011ApJS..197...12D}, below which there is no chance of finding inflated hot Jupiters.

Inflated planets are considered good candidates to study their atmospheric properties via transits \citep{Charbonneau_2005, 2009MNRAS.394..272S,winn_2008}. TOI-1789~b has a very low bulk density of $0.28^{+0.14}_{-0.12}$ g/cm$^3$. Assuming a hydrogen rich atmosphere and thereby a $\mu$ of 2.3, we calculate the scale height ($H=K_b T_{eq}/\mu g)$ \citep{2014prpl.conf..739M} of 852 km. The Transmission Spectroscopy Metric (TSM) for this target as noted in TFOP observing notes is 210.8. Based on the formulation from \cite{Kempton2018} and our derived parameters from this work, we re-calculate the TSM as 139.4. This places TOI-1789 in the top two quartiles for the sub-Jovian planets suitable for high-priority atmospheric characterization as noted by \cite{Kempton2018}. Despite this scale height and TSM, the expected amplitude of spectral features in transmission is only $\sim$ 0.016$\%$ \citep{2018haex.bookE.100K} due to the large radius of the host star. Thus, TOI-1789 might not be the ideal target for ground-based transmission spectroscopy studies but can be a suitable candidate for future JWST\footnote{\url{https://jwst.nasa.gov/science.html}} or ARIEL studies \citep{2016SPIE.9904E..1XT}. 

Due to its unique position in the evolutionary state and a relatively bright magnitude, TOI-1789 is a good choice to study the  Rossiter-McLaughlin (R-M) effect. Most hot Jupiters are aligned in their orbits with the spin angle of their host stars. However, some of these can also be misaligned \citep{2012ApJ...757...18A}. R-M effect \citep{1924ApJ....60...15R, 1924ApJ....60...22M} can be used to study this projected stellar obliquity of planets. Based on the larger stellar radius and its relatively high v~$\sin{i}$, RM semi-amplitude could be between 2.5~m~s$^{-1}$ and 16~m~s$^{-1}$ for the projected spin-orbit angle ($\lambda$) between $\ang{0}$ and $\ang{90}$, respectively \citep{rm}. Depending on the projected spin-orbit angle, this can be observed with any moderate-sized telescope (2.5 - 4 m aperture) with a precise RV instrument.

\section{Summary}\label{sec:summary}
We have presented the discovery and characterization of a transiting hot Jupiter, TOI-1789~b, from TESS photometry data, follow-up ground based-photometry data from PRL, along with precise RV data from PARAS and TCES spectrographs. Based on the global modelling of these photometry and RV datasets, the host star has a mass of $1.507^{+0.059}_{-0.14}$ \(M_\odot\), radius of $2.168^{+0.036}_{-0.034}$ \(R_\odot\), metallicity of $0.373^{+0.071}_{-0.086}$ dex, effective temperature of $5991\pm55$~K, and surface gravity ($\log{g_*}$) of $3.943^{+0.023}_{-0.043}$ that {indicate} it is a slightly evolved, metal-rich, and late F-type star.
 The planetary parameters for TOI-1789~b are as follows: \mpp: $0.70\pm0.16$ \mj, \rp: $1.44^{+0.24}_{-0.14}$ \rj, $\rho_P$: $0.28^{+0.14}_{-0.12}$ g cm$^{-3}$, and $T_{\rm eq}$: $1927^{+27}_{-17}$~K. At a 4-$\sigma$ significance on the derived mass and radius for TOI-1789~b, we find the radius of TOI-1789~b to be inflated for its planetary mass. TOI-1789~b could have already been circularized as seen from our results. This is in agreement with the calculated circularization timescale of 0.08 Gyr, which is less than the estimated age of the system. There are only 8 exoplanets (including TOI-1789~b) discovered as of now, which are hosted by either similar or more evolved stars than TOI-1789 and orbiting in very close-proximity to their host star (a $\leq$ 0.05 AU). Despite the rarity of hot Jupiters across slightly evolved stars in close-in orbits, TOI-1789~b is completely "non-anomalous," satisfying most of the evolutionary models at place. The detection of similar types of exoplanets will help us to improve our understanding of the distribution, formation, and migration of hot Jupiters around evolving stars.

\section*{Acknowledgements}

We acknowledge the PRL-DOS (Department of Space, Government of India) and the director of PRL for supporting the PARAS spectrograph funding for the exoplanet discovery project and the research grant for AK. PC acknowledges the generous support from Deutsche Forschungsgemeinschaft (DFG) of the grant HA3279/11-1. We are grateful for the generous support by Th\"uringer Ministerium f\"ur Wirtschaft, Wissenschaft und Digitale Gesellschaft. CMP and MF gratefully acknowledge the support of the Swedish National Space Agency (DNR 65/19-P). We acknowledge the help from Kapil Kumar, Vishal Shah, and all the Mount-Abu and TLS observatory staff for their assistance during the observations. We are grateful to the user support group of the Alfred Jensch telescope at Tautenburg, Germany. AC is grateful to Suvrath Mahadevan and Arpita Roy from Pennsylvania University, USA, for their tremendous efforts in developing the PARAS data pipeline in 2014. We thank Mathias Zechmeister and Jana Koehler for their help in the ongoing development of the VIPER pipeline. Some of the observations in the paper made use of the High-Resolution Imaging instrument ‘Alopeke obtained under Gemini LLP Proposal Number: GN/S-2021A-LP-105. ‘Alopeke was funded by the NASA Exoplanet Exploration Program and built at the NASA Ames Research Center by Steve B. Howell, Nic Scott, Elliott P. Horch, and Emmett Quigley. Alopeke was mounted on the Gemini North (and/or South) telescope of the international Gemini Observatory, a program of NSF’s OIR Lab, which is managed by the Association of Universities for Research in Astronomy (AURA) under a cooperative agreement with the National Science Foundation. On behalf of the Gemini partnership: the National Science Foundation (United States), National Research Council (Canada), Agencia Nacional de Investigación y Desarrollo (Chile), Ministerio de Ciencia, Tecnología e Innovación (Argentina), Ministério da Ciência, Tecnologia, Inovações e Comunicações (Brazil), and Korea Astronomy and Space Science Institute (Republic of Korea). This research has made use of the SIMBAD database \citep{simbad}, operated at CDS, Strasbourg, France. This research has made use of the Exoplanet Follow-up Observation Program website, which is operated by the California Institute of Technology, under contract with the National Aeronautics and Space Administration under the Exoplanet Exploration Program. This paper includes data collected with the TESS mission, obtained from the MAST data archive at the Space Telescope Science Institute (STScI). Funding for the TESS mission is provided by the NASA Explorer Program. STScI is operated by the Association of Universities for Research in Astronomy, Inc., under NASA contract NAS 5–26555. This work has made use of the ASAS3, SuperWASP, Keck archives, and the Transiting ExoPlanet catalogue (TEPcat) database.
The following software were used in research for this paper: AstroImageJ \citep{2017AJ....153...77C},  PARAS$\_$PIPELINE \citep{Chakraborty_2014}, EXOFASTv2 \citep{exofastv2}, \texttt{george} kernel \citep{Ambikasaran2015}, \texttt{juliet} \citep{juliet}, Astropy \citep{astropy}, \texttt{lightkurve} \citep{lightkurve}, {VIPER \citep{Zechmeister2021}} and PyAstronomy \citep{pya}.

\section*{Data Availability}
The TESS photometry and the high-resolution speckle imaging data underlying this article are available at the Mikulski Archive for Space Telescopes (MAST) ({\url{https://mast.stsci.edu/}}) and the ExoFOP-$TESS$ webpage (\url{https://exofop.ipac.caltech.edu/tess/target.php?id=172518755}), respectively. The radial velocity measurements are provided in the Table~\ref{tab:rv_table}. All other data underlying this article will be shared on reasonable request to the corresponding author.

\bibliographystyle{mnras}
\bibliography{references}
\renewcommand{\thefootnote}{\fnsymbol{footnote}}
\newpage
\onecolumn
\begin{center}
\begin{longtable}[c]{ccccccc}
    \caption{RV measurements for TOI-1789. The BJD$_{TDB}$ is mentioned in the first column followed by their corresponding relative RVs and error in RVs in the second and third column respectively. The fourth and fifth columns correspond to bisector velocity span and its respective errors. The sixth column represents the exposure time and the last column shows the instruments used for observations.}
	\label{tab:rv_table}\\ \hline
	    \noalign{\smallskip}
		
		BJD$_{TDB}$ & Relative-RV & $\sigma$-RV & BIS & $\sigma$-BIS & EXP-TIME & Instrument\\
		\vspace{0.1cm}\\
		Days & m s$^{-1}$ & m s$^{-1}$ & m s$^{-1}$ & m s$^{-1}$ & sec \\
		\noalign{\smallskip}
		\hline
		\noalign{\smallskip}
		
		2459203.519645 & -11.91 & 30.00 & 197.79 & 168.29 & 1800 & PARAS\footnotemark[1] \\
		2459236.471110 &  -91.76 & 28.13 & 267.54 & 64.79 & 1800 & PARAS\footnotemark[2] \\ 
		2459236.495532 & -106.14 & 31.24 & 307.58 & 58.14 & 1800 & PARAS\footnotemark[2]\\
        2459264.241224 & -34.16 & 26.40 & 972.78 & 93.32 & 1800 & PARAS\footnotemark[2]\\
        2459264.426243 & -53.28 & 38.74 &  -83.81 & 79.54 & 1800 & PARAS\footnotemark[2]\\
        2459266.309799 & -98.94 & 46.13 & 1751.67 & 220.13 & 1800 & PARAS\footnotemark[2]\\
        2459287.283471 & -44.48  & 20.80 & -551.95 & 30.95 & 1800 & PARAS\footnotemark[2]\\
        2459289.186833 & -125.73 & 46.89 & -982.45 & 153.12 & 1800 & PARAS\footnotemark[2]\\
        2459289.258414 & -100.47 & 23.03 & -100.73 & 87.37 & 1800 & PARAS\footnotemark[2]\\
        2459289.284072 & -87.48  & 22.52 & 0.70 & 36.28 & 1800 & PARAS\footnotemark[2]\\
        2459289.324973 & -134.37 & 23.88 & -415.21 & 28.96 & 1800 & PARAS\footnotemark[2]\\
        2459290.195198 & -31.61  & 25.76 & 175.88 & 36.30 & 1800 & PARAS\footnotemark[2]\\
        2459292.327774 & -26.77  & 31.32 & 86.37 & 59.11 & 1800 & PARAS\footnotemark[2]\\
        2459293.265260 & -38.99  & 25.86 & -52.63 & 26.21 & 1800 & PARAS\footnotemark[2]\\
        2459293.288928 &  16.25  & 28.94 & -80.99 & 101.90 & 1800 & PARAS\footnotemark[2]\\
        2459293.336772 & -68.44 & 33.08  & -79.96 & 90.14 & 1800 & PARAS\footnotemark[2]\\
        2459268.367086 &	55.72 &	30.32 & - & - & 1800 & TCES\\
        2459269.344931 &	4.01 &	28.88 & - & - & 1800 & TCES\\
        2459270.328739 &	211.94 &	31.26 & - & - & 1800 & TCES\\
        2459276.311054 &	142.06 &	37.66 & - & - & 1800 & TCES\\
        2459277.276355 &	133.96 &	37.06 & - & - & 1800 & TCES\\
        2459277.413716 &	115.03 &	32.70 & - & - & 1800 & TCES\\
        2459297.373577 &	42.26 &	14.10  & - & - & 1800 & TCES\\
        2459298.323521 & -25.34 &	14.19 & - & - & 1800 & TCES\\
        2459297.462606 &	136.68 &	39.96 & - & - & 1800 & TCES\\
        2459298.451947 &	61.38 &	20.11 & - & - & 1800 & TCES\\
        2459299.393539 &	292.36 &	31.03 & - & - & 1800 & TCES\\
        2459300.315441 &	151.34 &	29.86 & - & - & 1800 & TCES\\
        2459303.279352 &	179.31 &	34.34 & - & - & 1800 & TCES\\
        2459303.300500 &		187.65	 & 33.82 & - & - & 1800 & TCES\\
        2459303.436010 &	142.56 &	26.32 & - & - & 1800 & TCES\\
        2459303.457166 &	205.70 &	23.46 & - & - & 1800 & TCES\\
        2459305.274017	 &	128.12 &	24.60 & - & - & 1800 & TCES\\
        2459305.295176	&	101.73 &	29.09 & - & - & 1800 & TCES\\
        2459309.278728 &	281.19 &	30.41 & - & - & 1800 & TCES\\
        2459309.300456 &	226.44 &	27.72 & - & - & 1800 & TCES\\
        2459310.422304 &	113.38 &	38.13 & - & - & 1800 & TCES\\
        
		\hline
\end{longtable}
\begin{minipage}{.65\linewidth}
\footnotetext{\footnotemark[1] Spectra acquired simultaneously with ThAr HCL}
\footnotetext{\footnotemark[2] Spectra acquired simultaneously with UAr HCL}
\end{minipage}
\end{center}

\twocolumn

\newpage
\onecolumn
\begin{longtable}{lcccc}
\caption{Priors along with Median values and 68\% confidence interval for TOI-1789 from EXOFASTv2. The $\mathcal{N}$ and $\mathcal{U}$ represents the Gaussian and the Uniform priors, respectively.}
\label{result_exofast}\\
\hline
\noalign{\smallskip}
{Parameter} & {Units} & {Adopted Priors} & {Values}\\
\noalign{\smallskip}
\hline
\noalign{\smallskip}
\multicolumn{2}{l}{Stellar Parameters:}&\smallskip\\
~~~~$M_*$\dotfill &Mass (\msun)\dotfill &--&$1.507^{+0.059}_{-0.14}$\\
~~~~$R_*$\dotfill &Radius (\rsun)\dotfill&-- &$2.168^{+0.036}_{-0.034}$\\
~~~~$L_*$\dotfill &Luminosity (\lsun)\dotfill&-- &$5.45^{+0.15}_{-0.14}$\\
~~~~$\rho_*$\dotfill &Density (cgs)\dotfill&-- &$0.208^{+0.014}_{-0.020}$\\
~~~~$\log{g}$\dotfill &Surface gravity (cgs)\dotfill&-- &$3.943^{+0.023}_{-0.043}$\\
~~~~$T_{\rm eff}$\dotfill &Effective Temperature (K)\dotfill &$\mathcal{N}$(5894, 142)&$5991\pm55$\\
~~~~$[{\rm Fe/H}]$\dotfill &Metallicity (dex)\dotfill&$\mathcal{N}$(0.38, 0.1) &$0.373^{+0.071}_{-0.086}$\\
~~~~$Age$\dotfill &Age (Gyr)\dotfill&-- &$2.73^{+1.3}_{-0.51}$\\
~~~~$EEP$\dotfill &Equal Evolutionary Point \dotfill &--&$404^{+45}_{-14}$\\
~~~~$A_V$\dotfill &V-band extinction (mag)\dotfill&$\mathcal{U}$(0, 0.067) &$0.030^{+0.025}_{-0.021}$\\
~~~~$\sigma_{SED}$\dotfill &SED photometry error scaling \dotfill&-- &$2.78^{+0.74}_{-0.52}$\\
~~~~$v \sin{i}$\dotfill &Projected Rotational Velocity (km s$^{-1}$)\dotfill&-- &$7.0\pm0.5$\\
~~~~$\varpi$\dotfill &Parallax (mas)\dotfill& $\mathcal{N}$(4.4743, 0.0181) &$4.474\pm0.018$\\
~~~~$d$\dotfill &Distance (pc)\dotfill&-- &$223.53^{+0.91}_{-0.90}$\\
\smallskip\\\multicolumn{2}{l}{Planetary Parameters:}& &b\smallskip\\
~~~~$P$\dotfill &Period (days)\dotfill & --&$3.208664\pm0.000015$\\
~~~~$R_P$\dotfill &Radius (\rj)\dotfill&-- &$1.44^{+0.24}_{-0.14}$\\
~~~~$T_C$\dotfill &Time of conjunction (\bjdtdb)\dotfill&-- &$2458873.65374\pm0.00062$\\
~~~~$a$\dotfill &Semi-major axis (AU)\dotfill&-- &$0.04882^{+0.00063}_{-0.0016}$\\
~~~~$i$\dotfill &Inclination (Degrees)\dotfill&-- &$78.41^{+0.36}_{-0.58}$\\
~~~~$T_{eq}$\dotfill &Equilibrium temperature (K)\dotfill&-- &$1927^{+27}_{-17}$\\
~~~~$M_P$\dotfill &Mass (\mj)\dotfill&-- &$0.70\pm0.16$\\
~~~~$K$\dotfill &RV semi-amplitude (m/s)\dotfill &--&$72^{+17}_{-16}$\\
~~~~$logK$\dotfill &Log of RV semi-amplitude \dotfill&-- &$1.861^{+0.090}_{-0.11}$\\
~~~~$R_P/R_*$\dotfill &Radius of planet in stellar radii \dotfill&-- &$0.0680^{+0.011}_{-0.0061}$\\
~~~~$a/R_*$\dotfill &Semi-major axis in stellar radii \dotfill&-- &$4.83^{+0.11}_{-0.16}$\\
~~~~$\delta$\dotfill &Transit depth (fraction)\dotfill&-- &$0.00463^{+0.0016}_{-0.00080}$\\
~~~~$Depth$\dotfill &Flux decrement at mid transit \dotfill&-- &$0.00347^{+0.00020}_{-0.00018}$\\
~~~~$T_{14}$\dotfill &Total transit duration (days)\dotfill&-- &$0.0959\pm0.0018$\\
~~~~$b$\dotfill &Transit Impact parameter \dotfill&-- &$0.972^{+0.017}_{-0.011}$\\
~~~~$\rho_P$\dotfill &Density (cgs)\dotfill&-- &$0.28^{+0.14}_{-0.12}$\\
~~~~$logg_P$\dotfill &Surface gravity \dotfill&-- &$2.91^{+0.14}_{-0.18}$\\
~~~~$\fave$\dotfill &Incident Flux (\fluxcgs)\dotfill &--&$3.13^{+0.18}_{-0.11}$\\
~~~~$M_P\sin i$\dotfill &Minimum mass (\mj)\dotfill&-- &$0.68^{+0.16}_{-0.15}$\\
~~~~$M_P/M_*$\dotfill &Mass ratio \dotfill&-- &$0.00045\pm0.00010$\\

\noalign{\smallskip}
\noalign{\smallskip}\noalign{\smallskip}
\hline
\noalign{\smallskip}\multicolumn{2}{l}{Wavelength Parameters:}&R&TESS\smallskip\\
~~~~$u_{1}$\dotfill &linear limb-darkening coeff \dotfill &$0.356\pm0.035$&$0.263^{+0.046}_{-0.047}$\\
~~~~$u_{2}$\dotfill &quadratic limb-darkening coeff \dotfill &$0.309^{+0.035}_{-0.034}$&$0.284\pm0.047$\\
\smallskip\\\multicolumn{2}{l}{Telescope Parameters:}&PARAS&TCES\smallskip\\
~~~~$\gamma_{\rm rel}$\dotfill &Relative RV Offset (m/s)\dotfill &$-93\pm12$&$122^{+14}_{-13}$\\
\noalign{\smallskip}
~~~~$\sigma_J$\dotfill &RV Jitter (m/s)\dotfill &$26^{+14}_{-12}$&$51^{+14}_{-11}$\\
\noalign{\smallskip}
~~~~$\sigma_J^2$\dotfill &RV Jitter Variance \dotfill &$730^{+960}_{-510}$&$2700^{+1700}_{-1000}$\\
\smallskip\\\multicolumn{2}{l}{Transit Parameters:}&PRL ADR (R) & PRL TRI (R)&TESS (TESS)\smallskip\\
~~~~$\sigma^{2}$\dotfill &Added Variance \dotfill &$0.00000252^{+0.00000025}_{-0.00000023}$&$0.00000015^{+0.00000018}_{-0.00000016}$&$0.0000000064^{+0.0000000018}_{-0.0000000017}$\\
\noalign{\smallskip}
~~~~$F_0$\dotfill &Baseline flux \dotfill &$1.00006\pm0.00010$&$1.000509^{+0.000100}_{-0.000099}$&$0.9999953\pm0.0000061$\\
\noalign{\smallskip}
\hline
\end{longtable}

\newpage
\appendix

    \section{The corner plot showing the covariances for all the fitted parameters for the TOI-1789 global-fit}
\begin{figure}
\centering
	\includegraphics[scale=2]{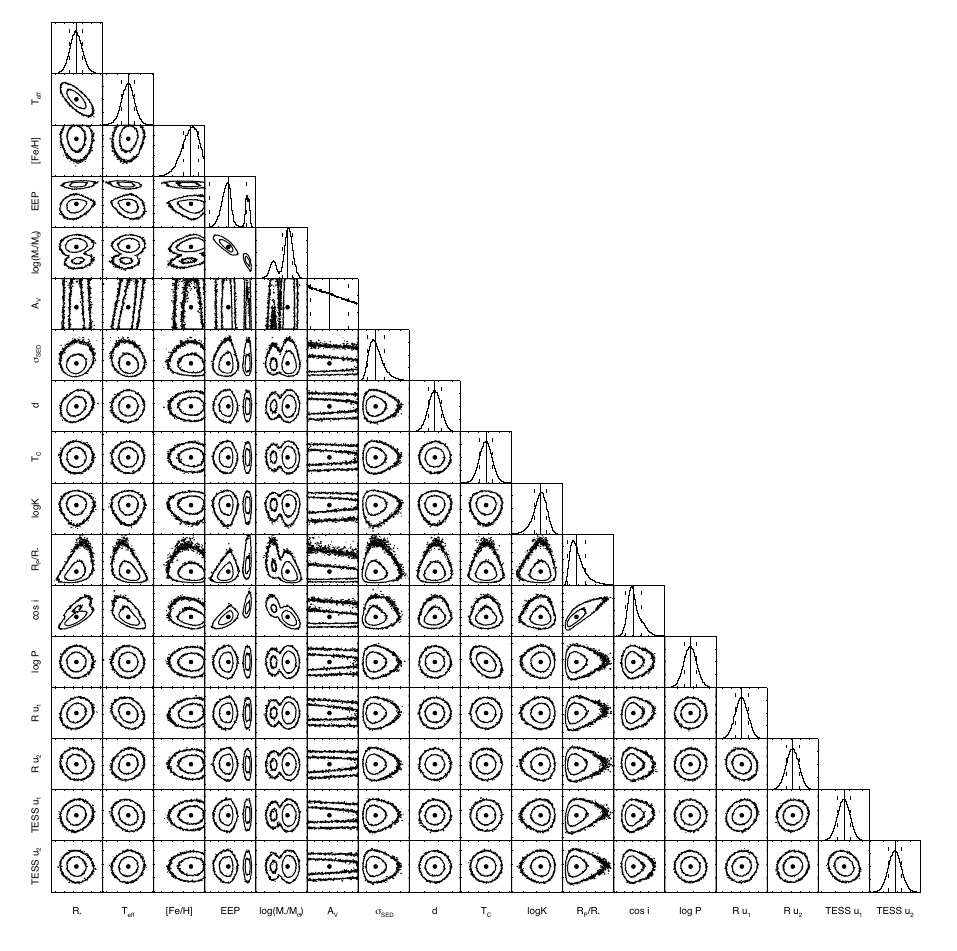}
	\caption{}
    \label{fig:corner}
    
\end{figure}

\label{lastpage}
\end{document}